\documentclass[conference]{IEEEtran}
\ifCLASSINFOpdf
\else
\fi
\usepackage[subpreambles=true]{standalone}
\usepackage[utf8]{inputenc}
\usepackage{makecell}
\usepackage{import}
\usepackage{graphicx}
\usepackage{seqsplit}
\usepackage{amssymb}
\usepackage{url}
\usepackage{amsmath}
\usepackage{booktabs}
\usepackage{tabularx}
\usepackage{multicol}
\usepackage{multirow}
\usepackage{xcolor}
\usepackage{algorithm}
\usepackage[noend]{algpseudocode}
\usepackage{rotating}
\usepackage{amssymb}
\usepackage{enumitem}
\usepackage{pifont}
\usepackage{xspace}
\usepackage{longtable}
\usepackage{subcaption}
\graphicspath{ {./images/} }
\usepackage{comment}

\newcommand{\mulval}{MulVAL\xspace}

\usepackage{verbatim}
\usepackage[utf8]{inputenc}
\usepackage[english]{babel}
\usepackage{amsthm}
\theoremstyle{definition}
\newtheorem{phase}{Phase}[]
\theoremstyle{definition}
\newtheorem{example}{Example}[]

\usepackage{authblk}
\newcommand*{\affaddr}[1]{#1} % No op here. Customize it for different styles.
\newcommand*{\affmark}[1][*]{\textsuperscript{#1}}

\begin{document}
\title{An Automated, End-to-End Framework for Modeling Attacks From Vulnerability Descriptions}

\author{%
Hodaya Binyamini\affmark[1], Ron Bitton\affmark[1], Masaki Inokuchi\affmark[2], Tomohiko Yagyu\affmark[2], Yuval Elovici\affmark[1] and Asaf Shabtai\affmark[1]\\
\affaddr{\affmark[1]Dept. of Software and Information Systems Engineering, Ben-Gurion University of the Negev }\\
\affaddr{\affmark[2]NEC Corporation}\\
}

\maketitle

\begin{abstract}
Attack graphs are one of the main techniques used to automate the risk assessment process.
In order to derive a relevant attack graph, up-to-date information on known attack techniques should be represented as interaction rules.
Designing and creating new interaction rules is not a trivial task and currently  performed manually by security experts.
However, since the number of new security vulnerabilities and attack techniques continuously and rapidly grows, there is a need to frequently update the rule set of attack graph tools with new attack techniques to ensure that the set of interaction rules is always up-to-date.
We present a novel, end-to-end, automated framework for modeling new attack techniques from textual description of a security vulnerability.
Given a description of a security vulnerability, the proposed framework first extracts the relevant attack entities required to model the attack, completes missing information on the vulnerability, and derives a new interaction rule that models the attack; this new rule is integrated within \mulval attack graph tool. 
The proposed framework implements a novel pipeline that includes a dedicated cybersecurity linguistic model trained on the the NVD repository, a recurrent neural network model used for attack entity extraction, a  logistic regression model used for completing the missing information, and a novel machine learning-based approach for automatically modeling the attacks as \mulval's interaction rule.
We evaluated the performance of each of the individual algorithms, as well as the complete framework and demonstrated its effectiveness.

\end{abstract}


% \fi
%%\IEEEpeerreviewmaketitle

\section{\label{sec:intro}Introduction}
Cybersecurity risk assessment is an essential activity that enables system stakeholders to assess the risks to their system and select suitable countermeasures~\cite{landoll2005security,ou2005mulval,stan2019extending}.
A traditional cybersecurity risk assessment procedure consists of the following steps: (1) identify system assets, (2) enumerate the threats to which those assets are exposed, (3) apply network mapping tools to derive the network topology, (4) apply a vulnerability scanner to reveal existing security vulnerabilities in system components, and (5) derive the attack surface of the system based on the information collected~\cite{albanese2014manipulating}. 
The attack surface represents the possible attack paths an attacker can take to compromise an asset, and thus it can be used to quantify the overall risk of the system. 
Based on the attack surface, an optimal mitigation strategy can be selected to eliminate the most critical attack paths.

Since modern environments are dynamic and continuously changing, and new attack techniques are constantly introduced by attackers, the attack surface of such environments also changes; therefore, risk assessment must be performed \textit{automatically and continuously}.
The successful implementation of an automated risk assessment process relies on the ability to automate the processes of network mapping (using Nmap~\cite{nmap}), vulnerability discovery (using tools such as Nessus~\cite{Nessuscite} or OpenVAS~\cite{developers2012open}), penetration testing (using advanced frameworks such as DeepExploit~\cite{DeepExploit} or Autosploit~\cite{moscovich2020autosploit}), and finally assessment, which includes three main tasks: deriving the attack surface, quantifying the risk, and identifying the optimal mitigation strategy that minimizes the risk.

Attack graphs are one of the main techniques used to perform the assessment process~\cite{inokuchi2019design}.
\mulval~\cite{ou2005mulval,ou2006scalable} was the first attack graph tool providing automatic end-to-end attack graph generation and analysis.
Specifically, \mulval can be used to derive the attack surface and quantify the risk of the system; based on the attack surface generated, various methods can be applied in order to automatically find the optimal mitigation strategy~\cite{stan2019heuristic,speicher2018formally,speicher2018stackelberg,kordy2018on,kordy2017how,fila2020exploiting}.

Despite recent research and improvements to attack graphs, there is one major challenge outstanding.
In order to derive a relevant attack graph, up-to-date information on known vulnerabilities should be available and represented. 
In attack graph tools (including \mulval) the vulnerabilities are represented using \textit{interaction rules} (e.g., the preconditions required for an attacker to execute code on a vulnerable host, the consequence of the attack) and \textit{facts} that specify the attacker and system state (e.g., the attacker’s initial location and goal, host and network configurations, and existing vulnerabilities).

\begin{figure}[b!]
    \centering
    \includegraphics[width=0.45\textwidth]{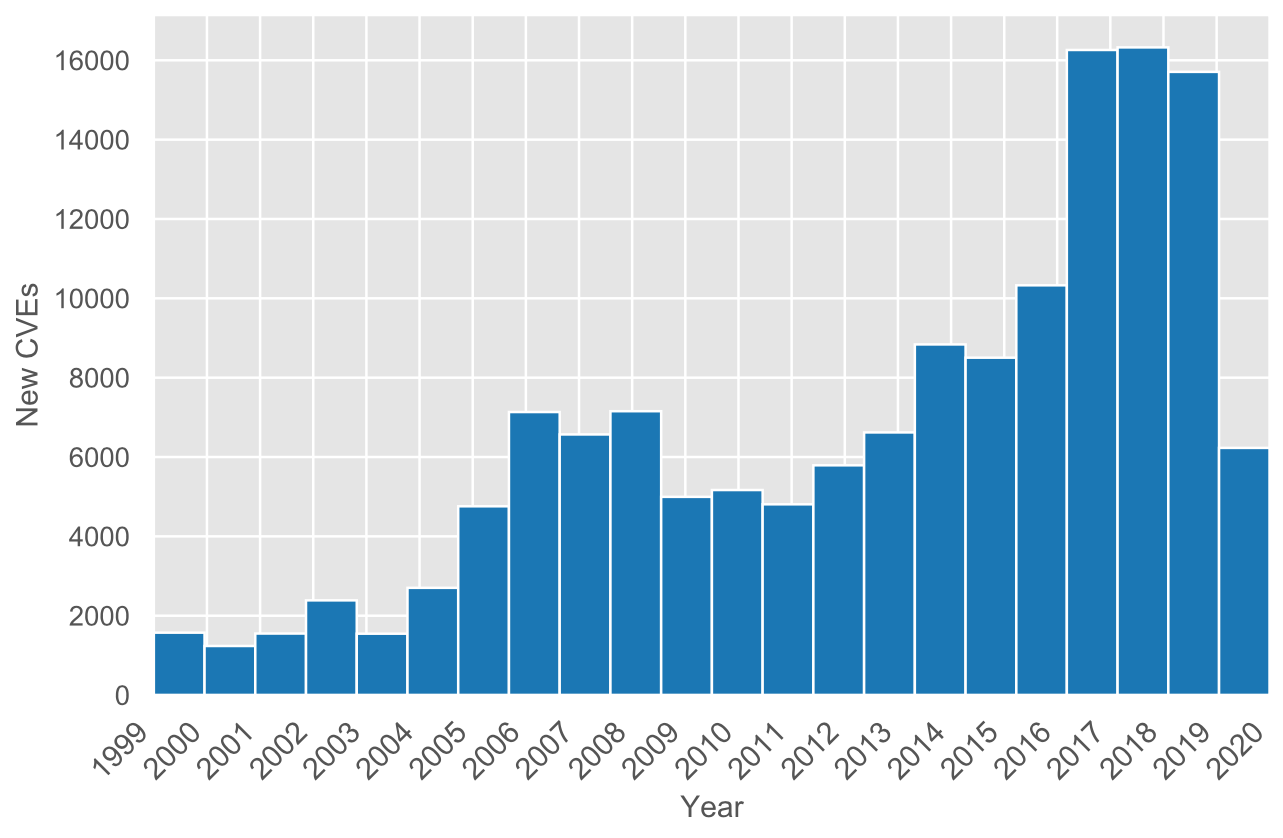}
    \caption{Vulnerabilities discovered from 1999 to 2020 according to the NVD repository.}
    \label{fig:NVD_inc}
\end{figure}

While facts can be derived automatically using tools such as Nmap, Nessus, and OpenVAS, designing and creating new interaction rules is not a trivial task and must be performed manually by security experts.
Because the attack landscape continuously and rapidly changes with new security vulnerabilities and attack techniques (see Figure~\ref{fig:NVD_inc}), there is a need to frequently update the rule set of attack graph tools with new attack techniques.
This research is aimed at developing an end-to-end, automatic framework for representing new vulnerabilities in attack graphs (specifically \mulval), thus ensuring that the set of interaction rules is always up-to-date. 

The development of an automated framework capable of expanding the set of interaction rules by adding new vulnerabilities and attack techniques must address the following three main challenges:
\begin{description}[style=unboxed,leftmargin=0cm]
\item[Automatically analyzing security vulnerabilities.]
When modeling a new attack technique, there is a need to specify both the attack's preconditions and consequence.
An attack's preconditions include the state required by the system for successful exploitation of the vulnerability (e.g., the vulnerable application should be running on the system), the context required for successful exploitation of the vulnerability (e.g., the attacker must have physical access to the vulnerable system), and the technique used by the attacker (e.g., the attacker sends a long message).
An attack's consequence includes the impact of the attack (e.g., executing code).
Although, this information is part of the Common Vulnerabilities and Exposures (CVE) standard~\cite{mann1999towards} (which defines the basic attributes of publicly known cybersecurity vulnerabilities), it is often written in free text;
while some of the attributes are structured and can be analyzed easily without human intervention, critical parts of the information can appear in natural language within the CVE description entry, thus making it more difficult to analyze the data automatically.

\item[Handling partial information.]
In many cases, only partial information about the vulnerability is provided~\cite{zhang2011empirical}; consequently, 
only partial information about the vulnerability is considered in the risk assessment procedure.

\item[Formulating the interaction rule.]
To create an interaction rule, there is a need to associate the attack entities (extracted from the description of the security vulnerability) with the predefined predicates.
Since attack entities are written in free text, the same preconditions/consequences can be appear in different wording.
As a result, mapping the preconditions/consequences to predicates is not a trivial task.
After mapping the preconditions/consequences to its predicates, the relationships between the predicates should be defined.
This is done by connecting the parameters of the various predicates.
Since the same parameter can be represented differently by different predicates, connecting the parameters correctly is also not a trivial task.
However, this task is crucial, since it defines the semantics of the interaction rule.
\end{description}
Previous works in this domain have focused on the first challenge, i.e., utilizing natural language processing (NLP) techniques to extract attack entities from descriptions of security vulnerabilities.
The vast majority of these methods are based on supervised machine learning (utilizing hand-crafted features)~\cite{lal2013information, weerawardhana2014automated, bridges2013automatic, aksu2018automated, jones2015towards} or rule-based approaches~\cite{jones2015towards,jing2017augmenting, weerawardhana2014automated, aksu2018automated}.
These methods, however, have several limitations.
A supervised machine learning approach cannot utilize the unlabeled data generally available (e.g., CVE repository).
Rule-based approaches do not consider the semantics of words, thus providing a very narrow solution that is difficult to generalize.
Although these methods were very popular in the past, state-of-the-art methods (such as Word2Vec, ELMo, and BERT) for NLP, which utilize the unlabeled data to construct a linguistic model, have been shown to be effective for improving many NLP tasks, including entity extraction~\cite{devlin2018bert}.
Recent research utilized linguistic models for extracting attack entities~\cite{li2019self, simran2019deep, kim2020automatic, gasmi2019cold}.
These methods, however, are solely based on pretrained linguistic models, without any fine-tuning.
Since the cybersecurity domain, and more specifically, attack descriptions, include specific terminology and linguistic semantics, the pretrained models available (which were trained on a generic corpus of data, such as English Wikipedia~\cite{zhu2015aligning}) are less suitable.

In this paper, we present a novel, end-to-end, automated framework for modeling new attack techniques and integrating them into the risk assessment process.
Given a description of a security vulnerability, the proposed framework \textit{(i)} extracts the relevant attack entities required to model the attack, \textit{(ii)} completes missing information on the vulnerability, \textit{(iii)} associates the attack entities to the predefined predicates, and \textit{(iv)} defines the relationships between those predicates, resulting in a new interaction rule that models the new attack technique. 

Within this framework, we implemented a novel pipeline that includes the following machine learning models which interact with one another:
\textit{1)} a dedicated cybersecurity linguistic model trained on 5.8M words from 146K vulnerability descriptions (from the NVD repository); \textit{2)} a recurrent neural network (BLSTM) which is used for attack entity extraction -- this network was trained on a unique dataset, created by us, of 20K labeled words from 650 vulnerability descriptions; \textit{3)} a  dedicated clustering models used for associating the attack entities to the predefined predicates; \textit{4)} a logistic regression model used for completing the missing information -- this model was trained on 40K vulnerability descriptions (which exist in the NVD repository); and \textit{5)}
an imputation model based on the $k$-nearest neighbors used to define the relationships between the predicates --  this model was trained on 200 rules that exist in \mulval's default interaction rule file.
We evaluated the proposed framework (which is based on a linguistic model trained on cybersecurity related content) and compared it with previous methods used for attack entity extraction.
The results show that the proposed method significantly outperforms existing methods~\cite{lal2013information, weerawardhana2014automated}.

In summary, the contributions of this paper are as follows:
\begin{itemize}
    \item An end-to-end framework for automatically modeling new attack techniques and integrating them into the risk assessment process. 
    \item A dedicated cybersecurity linguistic model trained on 5.8M words from 146K vulnerability descriptions (from the NVD repository); this model can be used for any downstream NLP task in the cybersecurity domain.
    \item An entity recognition model that can be used to extract attack entities from security vulnerabilities.
    This model is available as an online service for the security research community.\footnote{The link was removed to maintain the anonymity of the authors.}
    \item A labeled dataset of 20K labeled words (entities) from 650 vulnerability descriptions, which, to the best of our knowledge, is currently the largest dataset available.
\end{itemize}

\section{\label{sec:proposed-framewor}Overview of the Proposed Framework}
The proposed framework consists of four main phases which are presented in Figure~\ref{fig:framework} (in the Appendix): (1) derive a cybersecurity linguistic model, (2) extract attack entities, (3) complete the missing information, and (4) generate \mulval interaction rules.
In this section, we provide an overview of the phases and demonstrate the entire process using an example.

\begin{phase} \textbf{Derive a Cybersecurity Linguistic Model.}
In this phase, we utilize word embedding techniques (such as Word2Vec, ELMo, and BERT) in order to derive a language model for cybersecurity-related content.
The input for this phase is a repository of unstructured/semi-structured reports of cyberattacks -- in this research we used the NVD.
The output of this phase is a linguistic model.
Given a word or sentence written in free text, the linguistic model provides a numerical representation (vector) that preserves the semantic relations between words.
\end{phase}

\begin{phase} \textbf{Extract Attack Entities.}
In this phase, given a textual (structured or semi-structured) description of a vulnerability, we extract entities that are necessary for modeling the attack.
Examples of such entities are the attack vector, the means required by an attacker to exploit the vulnerability, the attack technique, the impact of the attack, the vulnerable platform, etc. 
The extraction of attack entities is performed in two steps.
First, the cybersecurity linguistic model derived in the previous phase is used to generate a numerical representation for the textual description of a vulnerability (i.e., the upstream task).
Second, given the numerical representation, a dedicated model (based on a bidirectional LSTM neural network) is used to extract attack entities (i.e., the downstream task).
\end{phase}

\begin{phase}\textbf{Complete Missing Information.}
In this phase, we utilize machine learning techniques to predict missing entities, based on similar attack reports.
The input for this phase is the list of entities extracted in the previous phase and the list of entities that are missing.
The output of this phase consists of the predicted values for the missing entities. 
\end{phase}

\begin{phase}\textbf{Generate \mulval Interaction Rules.}
In this phase, given the knowledge extracted about the attack, we generate the \mulval's interaction rules. 
This is done by utilizing machine learning techniques.
The inputs for this phase are: (1) the list of entities and the values extracted from the attack description, and (2) the completed values derived in the previous phase. 
The output of this phase is the \mulval interaction rule that models the attack. 
\end{phase}

\begin{figure*}[t!]
    \centering
    \includegraphics[width=1\textwidth]{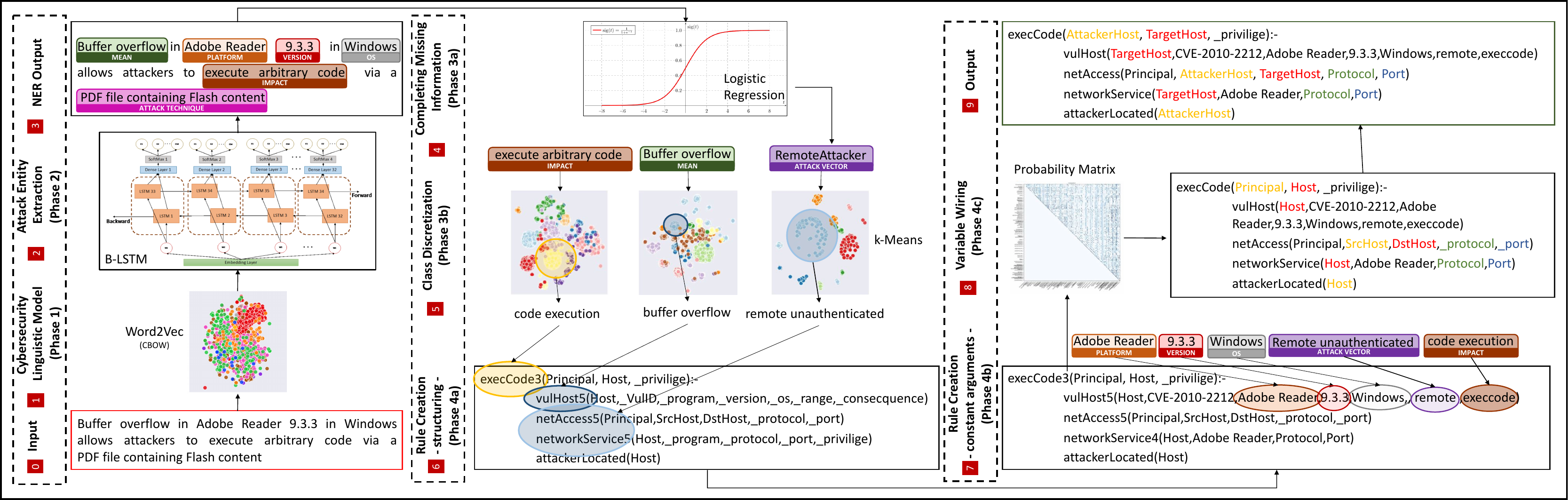}
    \caption{The process of generating MulVAL's interaction rule, given a description of security vulnerability.}
    \label{fig:cve-example}
\end{figure*}

\begin{example}
A concrete example for the four main phases of the proposed framework is presented in Figure~\ref{fig:cve-example}.
The input is a free text description of a security vulnerability in Adobe Reader that appears in the NVD (CVE-2010-2212).
First, we utilized the cybersecurity linguistic model to generate a numerical representation for each word in the description (Phase~1).
Those vectors are the input to the attack entity extraction algorithm, which extracts the entities that are necessary for modeling the attack (Phase~2). 
As can be seen, the algorithm identifies six different entities: the means used by the attacker to exploit the vulnerability (buffer overflow); the vulnerable platform (Adobe Reader); the vulnerable versions (9.0.0-9.3.3 and 8.0.0-8.2.3); the vulnerable operating systems (Windows and Mac OS X); the impact of the attack (execute arbitrary code or cause denial of service); and the attack technique (PDF file containing Flash content with a crafted tag).

Once the attack entities are extracted, we identify and complete the missing entities (Phase~3).
In this example, the attack vector, which is an extremely important property of the attack, is not mentioned within the description.
The proposed method was able to complete this missing value and identified that the vulnerability can be exploited remotely (i.e., remote attack vector).

Finally, given the attack entities, the proposed method generates a \mulval interaction rule that models the attack (Phase~4).
As can be observed, this rule consists of the following five preconditions: a target host (denoted as $Host$) running an Adobe Reader application (version 9.0.0-9.3.3 or 8.0.0-8.2.3 on Windows or Mac OS X); a host controlled by the attacker (denoted as $AttackerHost$); a remotely exploitable vulnerability in Adobe Reader (in the specified versions), which leads to a privilege escalation; and a network access from the attacker host to the target host.
Satisfying these preconditions allows the attacker to execute code on the target machine by exploiting the vulnerability.
We will elaborate on each of the phases in next sections.
\end{example}

\section{\label{sec:embedding}Cybersecurity Linguistic Model}
Linguistic models have been shown to be effective for improving many NLP tasks \cite{devlin2018bert,peters2018deep, mikolov2013efficient}.
These include generic tasks, such as question answering, named entity recognition, and sentiment analysis, as well as content-specific tasks, such as attack entity extraction from cybersecurity reports \cite{li2019self, simran2019deep, kim2020automatic, gasmi2019cold}.
The main advantages of linguistic models over traditional approaches are threefold:
First, linguistic models decouple the upstream task (i.e., learning general language representation) from the downstream tasks (e.g., sentiment analysis), thus enabling the linguistic model to be reused in different applications.
Second, linguistic models can utilize unlabeled data, which is widely available.
Third, linguistic models preserve the semantics of words.
For example, words with a similar meaning or words that appear in similar contexts will be close to each other within the latent representation.

When applying linguistic models in a new domain, two alternatives can be considered: (1) training a domain-specific linguistic model from scratch, or (2) utilizing an existing linguistic model (such as BERT~\cite{devlin2018bert}, GloVe~\cite{pennington2014glove}, or ELMo~\cite{peters2018deep}), which was already trained on a large corpus of data from another domain.
When using a pretrained model, the model can be applied without any changes or by fine-tuning the model with domain-specific data.
Intuitively, when a large corpus of domain-specific data is unavailable or when computational power is limited, training a model from scratch cannot be considered.
On the other hand, when the target domain has specific terminology and linguistic semantics, pretrained models often don't perform well.

Recent research on attack entity extraction have been based solely on pretrained linguistic models, without any fine-tuning~\cite{li2019self, simran2019deep, kim2020automatic, gasmi2019cold}.
Since the cybersecurity domain, and more specifically, attack descriptions include specific terminology and linguistic semantics, we believe that a linguistic model specifically trained on cybersecurity related content will outperform the pretrained models available (which are trained on a generic corpus of data, such as English Wikipedia \cite{zhu2015aligning}).
For this reason, we opted to develop a dedicated cybersecurity linguistic model.

\subsection{\label{subsec:dataset}Dataset}
\begin{figure}[tb!]
    \centering
    \includegraphics[width=0.5\textwidth]{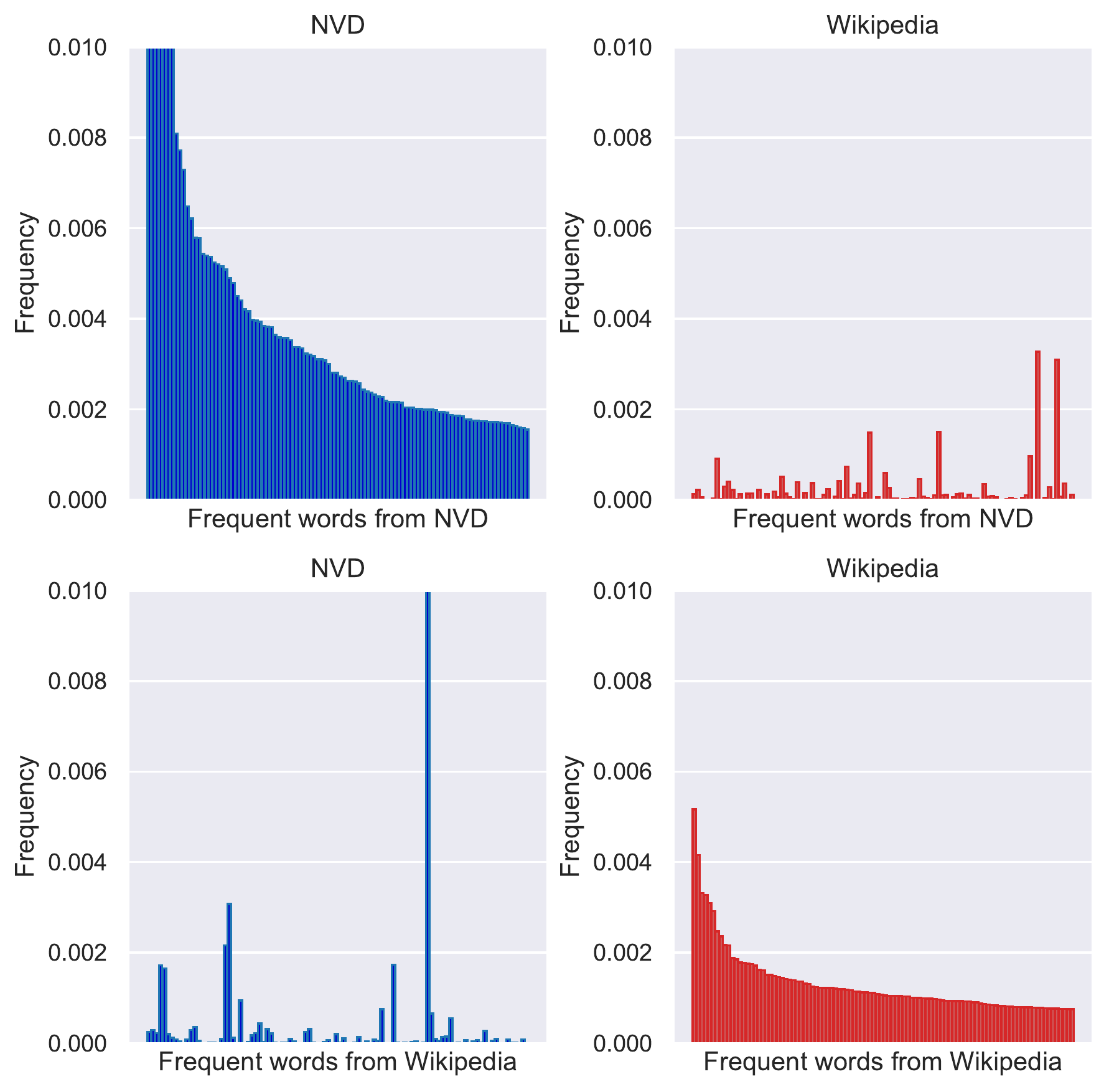}
    \caption{Presenting the distributions of the most frequently occurring words from the NVD and Wikipedia dataset.}
    \label{fig:words_dist}
\end{figure}

The dataset used to create our cybersecurity linguistic model consists of 146K vulnerability descriptions from the NVD (253K sentences and 5.8M words in total).
In order to support our decision for creating of a new linguistic model for cybersecurity, we compared frequently occurring words in the NVD dataset with frequently occurring words in Wikipedia (T-REx dataset~\cite{elsahar_2017}).
Specifically, we extracted the most frequently occurring words (after removing stop words) from the NVD and Wikipedia (the top 100 from each dataset) and explored their distribution. 
The results are presented in Figure~\ref{fig:words_dist}.
The top two graphs present the frequency of the 100 most frequent words in the NVD dataset, where graph (a) presents the words' distribution within the NVD dataset, and graph (b) presents their distribution within the Wikipedia dataset. 
Similarly, the graphs below present the frequency of the 100 most frequent words in the Wikipedia dataset, where graph (c) presents the words' distribution within the NVD dataset and graph (d) presents their distribution within the Wikipedia dataset. 
As can be seen, the distributions are quite different -- frequently occurring words in the NVD dataset are not so frequent in the Wikipedia dataset and vice versa.
For example, while the most common words in the Wikipedia dataset are: \textit{also, one, first, used, may, known, many, two, world, and united}, the most common words in the NVD dataset are: \textit{via, allows, remote, attackers, vulnerability, arbitrary, execute, service, code, and cause}.
This observation supports our main assumption that the cybersecurity domain has specific terminology and linguistic semantics, and therefore a linguistic model that is specifically trained on cybersecurity related content will outperform the pretrained models available, which are trained on a generic corpus of data, such as English Wikipedia.

\subsection{Word embedding algorithm}
State-of-the art algorithms for creating linguistic models include millions of parameters. 
In contrast to fine-tuning a pretrained model, training a linguistic model from scratch requires a large corpus of data.
For example, the language representation model, BERT~\cite{devlin2018bert}, consists of 340M trainable parameters and was trained on a very large corpus of 3B words.
In contrast, our dataset, which is based on vulnerability descriptions present in the NVD, consists of 5.8M words.
Furthermore, the computational power required to train models, such as BERT, is extremely high.
The BERT training was performed on 64 TPU chips for a period of four days~\cite{devlin2018bert}.
The estimated cost for this training (based on Google Cloud's price per hour for on-demand TPU v3) is about \$50K.
Hence, for our purposes, training complex and deep models like BERT form scratch cannot be considered, and therefore we opted to use a lightweight model.

The proposed linguistic model was created using Word2Vec, a lightweight language representation algorithm~\cite{mikolov2013efficient}.
In contrast to deep neural network architecture, such as BERT~\cite{devlin2018bert} and ELMo~\cite{peters2018deep}, Word2Vec utilizes a shallow neural network with only one hidden layer, thus dramatically reducing the number of trainable parameters, which directly affects the amount of data and computational power required to train a model.

Specifically, we tested two popular variants of Word2Vec: continuous bag-of-words (CBOW)~\cite{mikolov2013efficient} and skip-gram (SG)~\cite{mikolov2013efficient}.
The main difference between the two variants is that in CBOW the model tries to predict the current word given its surrounding contextual words, while in SG the model tries to predict the context (surroundings words) of a given word~\cite{mikolov2013efficient}.

\begin{description}[style=unboxed,leftmargin=0cm]
\item[Network architecture.]
In our implementation, the network architecture of the two models (CBOW and SG) is very similar: we set the vocabulary size to 10K words (which covers 93\% of the text in the NVD dataset), the window size to five words, and the embedding dimension size to 100 neurons.
As a result, the architectures of the models consist of a shallow neural network with three layers: an input layer with a size of 10K inputs (in one-hot vector encoding), a \textit{dense} hidden layer with 100 neurons, and a \textit{softmax} output layer with a size of 10K.
There are 2M trainable parameters, which is much smaller than deep neural network architectures.

\item[Optimization.]
We trained the models using backpropagation~\cite{rumelhart1985learning} for 300 epochs and stochastic gradient descent~\cite{sutskever2013importance}, where the learning rate was set at 0.0001.
\end{description}

\subsection{Evaluation}
\begin{description}[style=unboxed,leftmargin=0cm]
\item[Experimental setup.]
In order to evaluate the proposed cybersecurity linguistic models, we compared the Word2Vec models with state-of-the-art pretrained ELMo and BERT models.
For each model we tested the following two setups: using the pretrained model without any fine-tuning and using the pretrained model with fine-tuning.
The fine-tuning was performed by continuing the training of the pretrained models (for 100 epochs), using the dataset of attack descriptions from the NVD (as presented in Section~\ref{subsec:dataset}).
All of the experiments were performed on a single GPU of NVIDIA GeForce Titan X Pascal with 12GB.
 
\item[Evaluation measures.]
The comparison was performed based on the three common approaches for evaluating linguistic models: word semantic similarity, concept categorization, and downstream task performance~\cite{bakarov2018survey, peters2018deep, devlin2018bert}. 

\begin{itemize}
    \item \textit{Word semantic similarity.}
    This method is based on the rationale that the distance between words in the embedding space should reflect the semantic distance between these words.
    In order to evaluate the semantic similarity of words, we selected a list of words from the cybersecurity domain and let the different linguistic models find the words closest to these words in the embedding space.
    
    \item \textit{Concept categorization.}
    This method evaluates the clusters within the word embedding space.
    The rationale for using clustering is based on the assumption that words belonging to the same category will be close to each other within the embedding space.
    We evaluate the clusters with respect to categories from the cybersecurity domain.
    Specifically, we used attack entities, such as the attack vector, attack technique, mean, impact, and operating system, as categories.
    
    \item \textit{Downstream task performance.}
    This method evaluates the performance of the different linguistic models in the downstream task. 
    The rationale is that a better linguistic model will produce a better feature vector for the supervised machine learning algorithm, which will result in better performance in the downstream task. 
    In our context, the downstream task is attack entity recognition, which extracts the relevant entities of the attack from a given text that describes the attack. 
    In this evaluation, we tested the performance of the attack entity extraction algorithm when using different cybersecurity linguistic models.  
\end{itemize}

\item[Results.]
We present the results with respect to the first two evaluation measures, namely: word semantic similarity and concept categorization.
The results for the third evaluation measure i.e., the performance in the downstream task, are presented in Section~\ref{sec:attack-entity-recog}.
\begin{itemize}
    \item \textit{Word semantic similarity.}
    The results are presented in Figure~\ref{fig:word_semantic_similarity}.
    The words that we chose for this evaluation are: \textit{buffer, privilege, kernel, and windows}.
    As can be seen, the fact that the ELMo model used character embedding, influences the neighbors of each word, even if the meaning of the words is different.
    In the BERT model, we can see that the dataset used for pretraining affects the neighbors; for example, the neighborhood of \textit{windows} consists of the words \textit{doors, rooms, and window} which do not exist in the cybersecurity domain.
    In the graphs that describe the CBOW and SG models, we can see that all of the neighbors of each word have the same security-related meaning (and even the same entity).

    \begin{figure}[t!]
        \centering
        \includegraphics[width=0.47\textwidth]{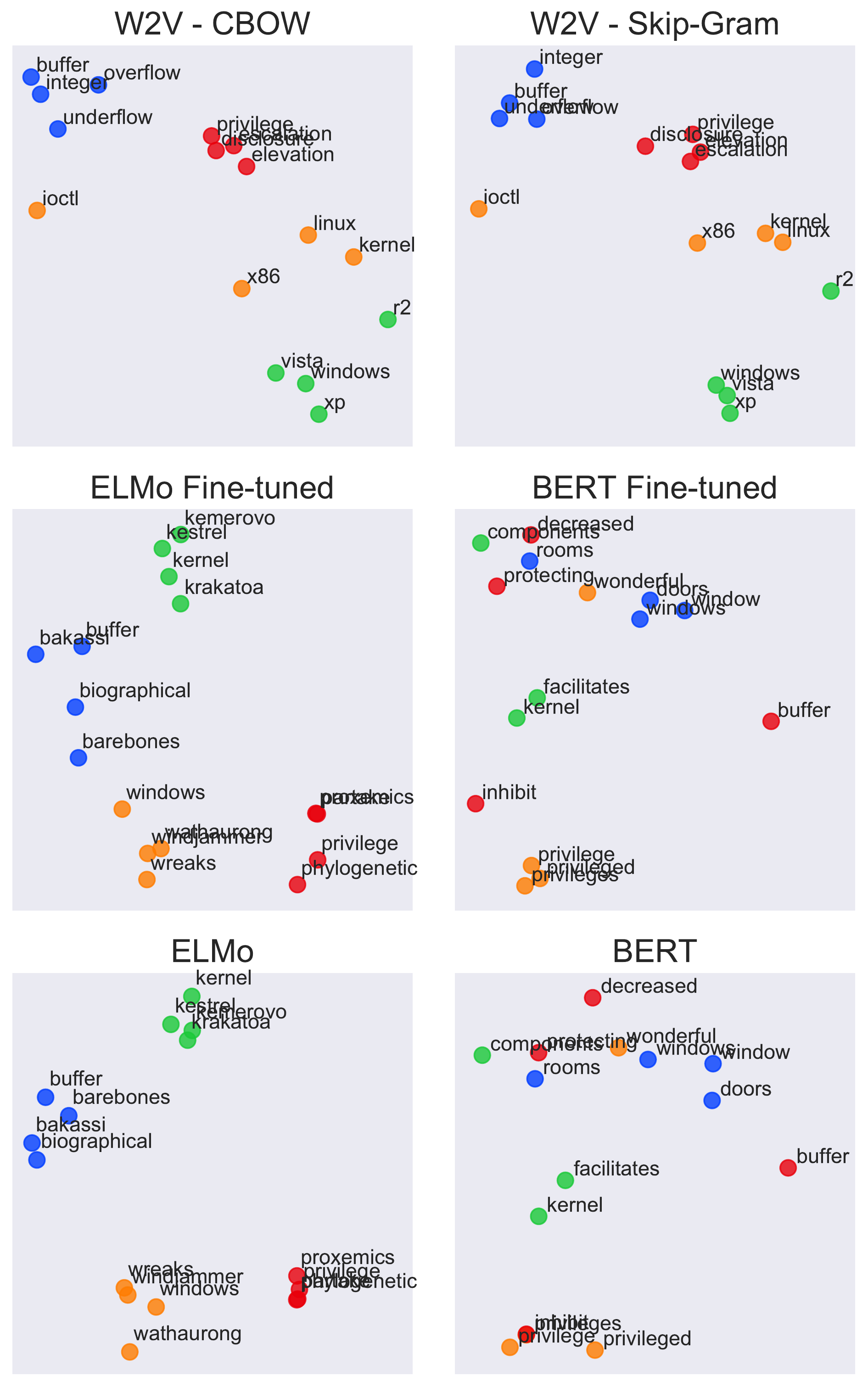}
        \caption{A visualization (using T-SNE) of the word embedding space of each linguistic model for selected words from the cybersecurity domain and their closest neighbours in the embedding space. 
        Each color in the graphs indicates a cluster of closest words in the embedding space.}
        \label{fig:word_semantic_similarity}
    \end{figure}
    
    \item \textit{Concept categorization.}
    The results are presented in Figure~\ref{fig:concept_categorization}.
    Each sub-figure visualizes the word embedding space of the different linguistic models (we used the T-SNE algorithm for dimensional reduction).
    As can be seen, the word embedding space of the pretrained BERT and ELMo models does not show clear clusters.
    Similar results were observed for the fine-tuned BERT and ELMo models.
    In contrast, the word embedding space of the proposed models (which were trained from scratch on cybersecurity related content) contains clear clusters that represent the different attack entities.
    
    \begin{figure}[t!]
        \centering
        \includegraphics[width=0.47\textwidth]{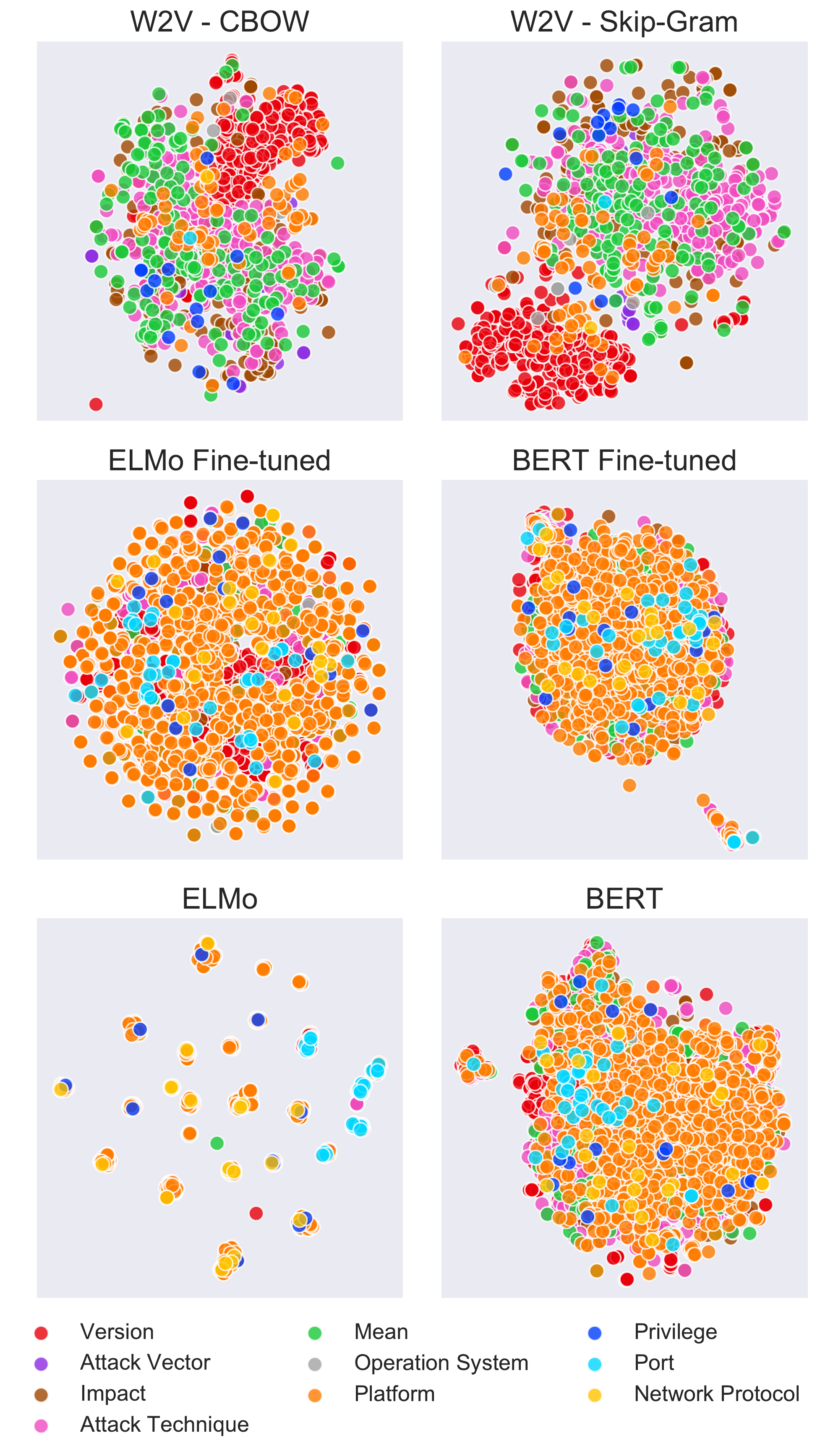}
        \caption{A visualization (using T-SNE) of the embedding space for each linguistic model, for words from the NVD corpus.
        Each color in the graphs indicates a different attack~entity.}
        \label{fig:concept_categorization}
    \end{figure}
\end{itemize}
\end{description}

\section{\label{sec:attack-entity-recog}Attack Entity Extraction}
Entity extraction, or named entity recognition (NER), is a common NLP task~\cite{sharnagat2014named}.
The main objective of this task is to classify unstructured text (written in natural language) into predefined categories.
Formally, given a sequence of words $S = (w_1,...,w_{n_w})$ and predefined categories $C = (c_1,...,c_{n_c})$, the output is essentially a sequence of categories (tags) $T = (t_1,...,t_{n_w} | t_i \in C)$ that maps each word to each category.

Entity extraction has been researched for more than two decades.
Early methods were based on handcrafted rules, lexicons, orthographic features, and ontologies~\cite{yadav2019survey}.
These methods were followed by supervised learning algorithms, such as the hidden Markov model~\cite{bikel1999algorithm,zhou2002named}, maximum entropy models~\cite{borthwick1999maximum,curran2003language}, the support vector machine~\cite{mcnamee2002entity}, and conditional random fields~\cite{mccallum2003early}.
Although these methods were very popular in the past, they have two main limitations.
First, these methods are based on manual feature engineering, which does not utilize the unlabeled data that is most available.
Second, these methods can only process a single data point and not entire sequences.
Therefore, they cannot be used to learn long-term dependencies.

Recent advances in NER utilize linguistic models for automatic feature extraction and recurrent neural networks (in particular, a special type of RNN termed the bidirectional long short-term memory network \cite{bengio1994learning}) for the entity extraction task \cite{yadav2019survey}.
Based on recent advances in NER, we developed a dedicated attack entity extraction algorithm.
This algorithm utilizes the cybersecurity linguistic model (described in Section~\ref{sec:embedding}), as well as a dedicated BLSTM neural network architecture, to extract multiple attack entities.

\subsection{Attack entities}
In this work, we consider the following attack entities:

\begin{description}[style=unboxed,leftmargin=0cm]
\item[Platform.]
Security vulnerabilities are usually associated with a specific product.
This attack entity is used to identify product names.
In this study, we consider two types of products: software products (e.g., Adobe Reader) and hardware products (e.g., Cisco Gateway).

\item[Version.]
Security vulnerabilities are often bound to some specific versions of a product. 
This attack entity is used to identify specific product versions (e.g., 9.x before 9.3.3 and 8.x before 8.2.3).

\item[Operating system.]
In some cases, vulnerability exploitation is bound to a specific operating system.
This attack entity is used to identify the names of operating systems (e.g., Windows or Linux).

\item[Network protocol.]
In some cases, security vulnerabilities exist within network protocols or can be exploited through network protocols.
This attack entity is used to identify network protocols (e.g., Telnet).

\item[Network port.]
In cases in which the vulnerability is bound to a network protocol, the default port is usually mentioned.
This attack entity is used to identify the network port (e.g.,~23).

\item[Attack means.]
This attack entity is used to identify the specific weakness exploited by the attacker (e.g., cross-site scripting, buffer overflow).  

\item[Attach technique.]
This attack entity is used to identify the specific technique used by the attacker to exploit the vulnerability (e.g., sending a crafted HTTP request).

\item[Attack impact.]
This attack entity is used to identify the specific consequences of executing the attack (e.g., execute code, denial of service, privilege escalation).

\item[Privilege.]
This attack entity is used to identify the privilege required by the attacker to exploit the vulnerability or the privilege granted to the attacker after exploiting the vulnerability (e.g., user privilege, root privilege). 

\item[Attack vector.]
This attack entity is used to identify the context required for successful  exploitation  of  the  vulnerability (e.g., the  attack requires the attacker to have physical access to the vulnerable system).

\end{description}

\subsection{Dataset}
Because a large dataset of security vulnerabilities with labeled attack entities described above wasn't availble at the time of this research, we created a labeled dataset consisting of 650 vulnerability descriptions from the NVD (800 sentences and 20K words in total).
The labeling was performed by 245 computer science students during an academic cybersecurity course, with each vulnerability labeled by at least two students.
We validated the labeling by comparing the labels assigned by the students; in cases in which there was a discrepancy, we selected the correct label.

\begin{figure}[h]
    \centering
    \includegraphics[width=0.47\textwidth]{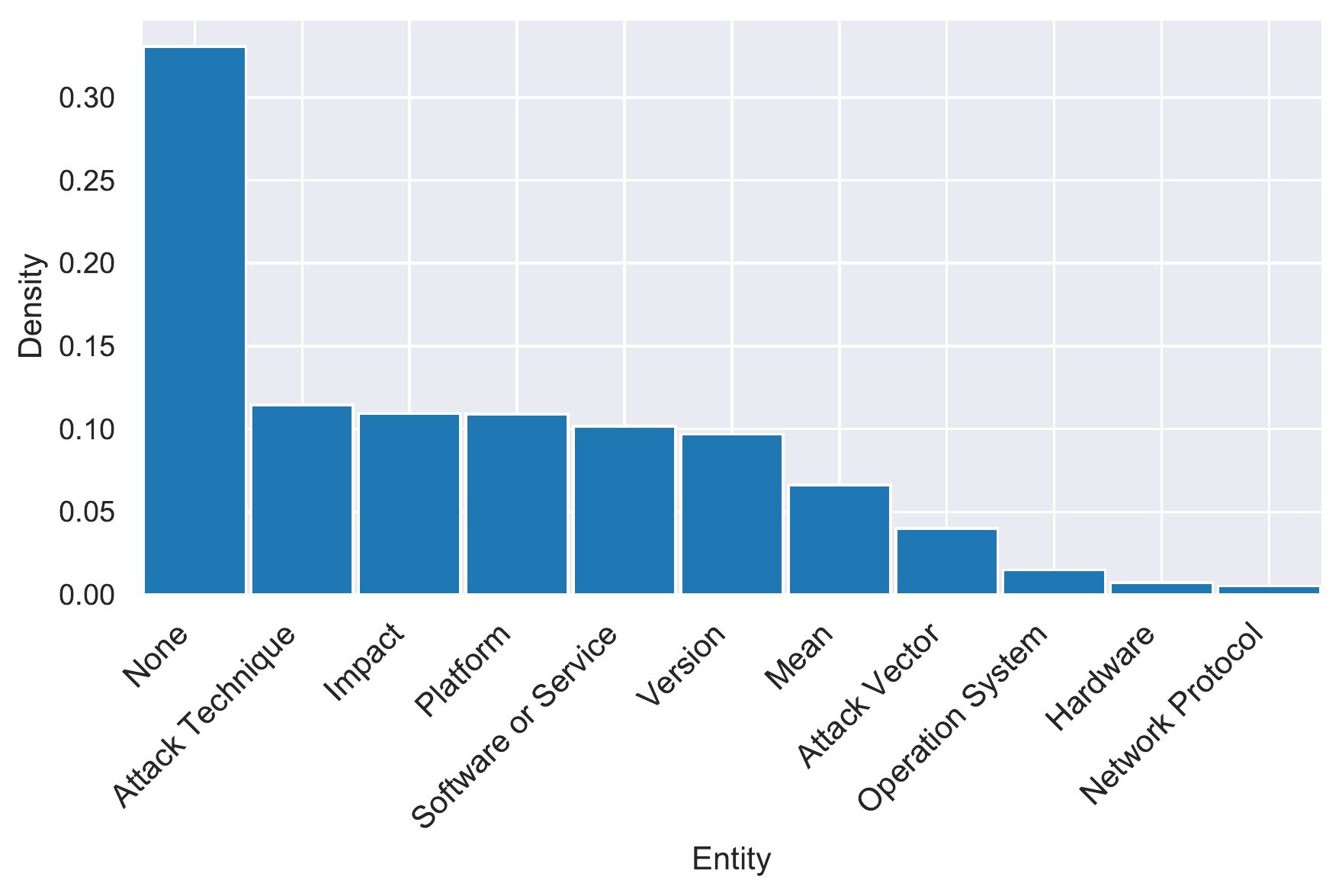}
    \caption{Histogram of the data entities.}
    \label{fig:data_dist}
\end{figure}

Descriptive statistics about the dataset are presented in Figure~\ref{fig:data_dist}.
As can be seen, the dataset is imbalanced, since a large portion of the words (about 30\%) do not match any entity.
In addition, some entities are less common (such as privilege, protocol, and port).

\subsection{Attack entity extraction algorithm}
\begin{description}[style=unboxed,leftmargin=0cm]
\item[Network architecture.]
The input to the attack entity extraction algorithm is a sequence of words, which are represented using the 
cybersecurity linguistic model (described in Section~\ref{sec:embedding}.
The output for each word is a distribution of the attack entities.
The architecture of the proposed neural network is presented in Figure~\ref{fig:ner}.
The network includes the following layers:
an input layer with $N$ neurons, each of which has $D$ inputs, where $N$ represents an upper bound for the length of a sentence, and $D$ represents the embedding dimension;
a forward and backward LSTM layer, each of which has $N$ LSTM units;
$N$ dense layers with $O$ neurons, each of which has $2D$ inputs, where $O$ represents the number of attack entities; and 
$N$ softmax output layers.   

\begin{figure}[tb!]
    \centering
    \includegraphics[width=0.5\textwidth]{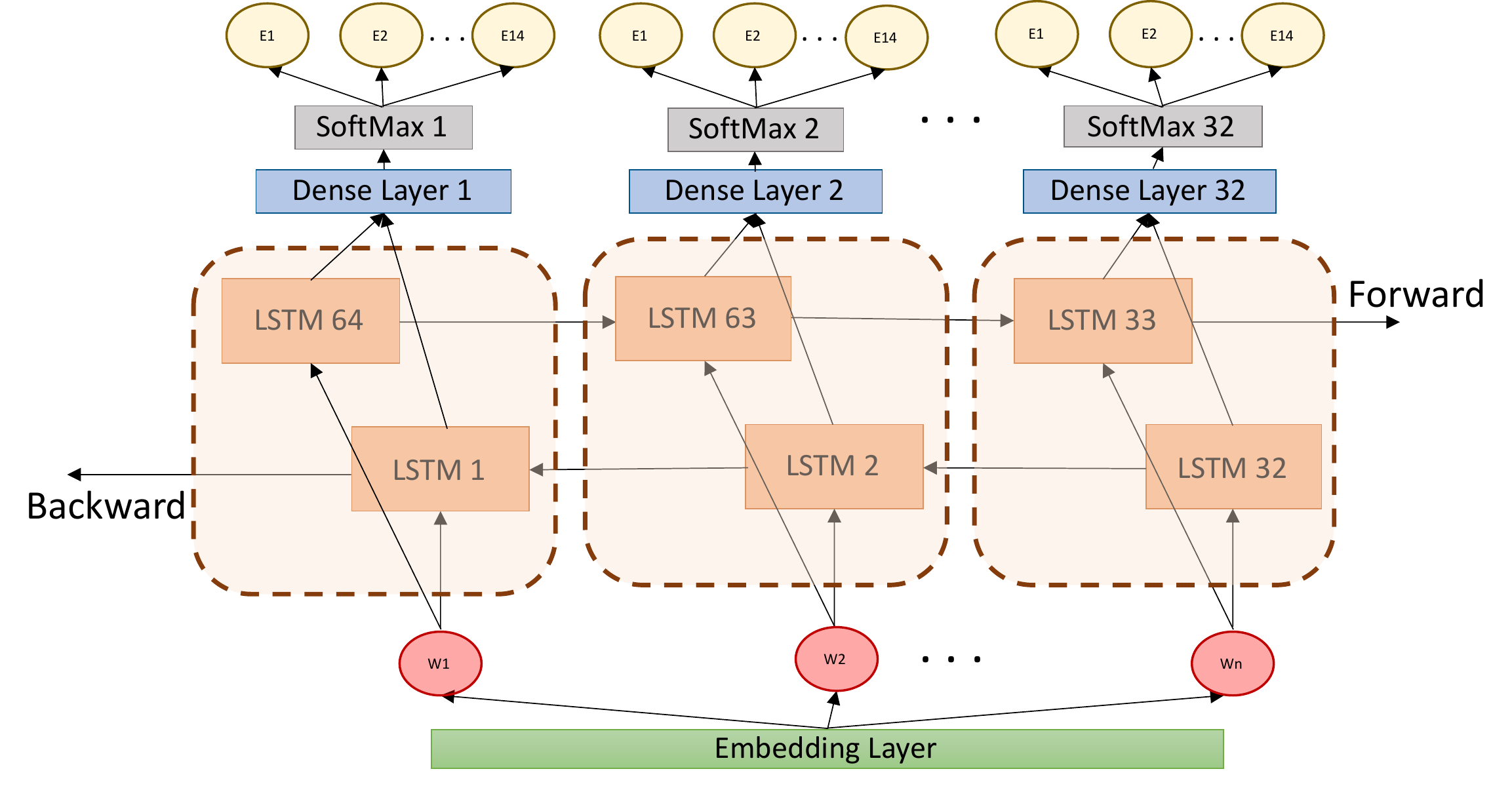}
    \caption{The proposed entity extraction algorithm.}
    \label{fig:ner}
\end{figure}

\item[Optimization.]
We trained the models using backpropagation through time~\cite{werbos1990backpropagation,  graves2013speech} for 100 epochs (and a batch size of 32), using stochastic gradient descent~\cite{sutskever2013importance}, where the learning rate was set at 0.01.
In addition, to cope with the imbalanced dataset, we used the following weighted cross-entropy loss function.
\begin{align}
        L &= - \sum_{i}w_{i}y_i\log(p_i) \quad
        w_{i}&= 
            \begin{cases}
                10,& \text{if } i \leq 10\\
                1,              & \text{otherwise}
            \end{cases}
\end{align}
where the first 10 classes are the attack entities, and the last class is the `none' entity.
\end{description}

\subsection{Evaluation}
\begin{description}[style=unboxed,leftmargin=0cm]
\item[Experimental setup.]
We tested the proposed NER architecture with the different linguistic models described in Section~\ref{sec:embedding}.
Specifically, we trained a different NER model for each of the following  linguistic models: Word2Vec models (CBOW and SG), pretrained ELMo and BERT models, and fine-tuned ELMo and BERT models.
Since the dimension of the embedding space is different for the various linguistic models examined, we set the $D$ parameter of the NER model in accordance with that dimension, i.e., for the Word2Vec, $D$ was set to 100; for the ELMo models, $D$ was set to 1024; and for the BERT models, $D$ was set to 768.
In addition, we compared the proposed NER model architecture to the architecture used in previous studies specifically focused on attack entity extraction.
These works were based on supervised machine learning algorithms (such as the conditional random field algorithm)~\cite{lal2013information, weerawardhana2014automated} and rule-based systems~\cite{weerawardhana2014automated}.
In order to evaluate the models, we used 5-fold cross-validation (such that each fold includes 130 randomly selected CVEs).
All of the experiments were performed on a single GPU of NVIDIA GeForce Titan X Pascal with 12GB.

\item[Evaluation measures.]
We used the F-score (F-measure)~\cite{tharwat2018classification}, which is the harmonic mean of the precision and the recall, as a measure for the accuracy of the models.
Since we are dealing with a multiclass classification problem, we used a one versus all setup to calculate F-scores for specific attack entities and the micro/macro average for the average performance.

\item[Results.]
Several interesting observations can be made regarding the results which are presented in Table~\ref{table-ner-results}.
First, the proposed NER model yields superior results when using the cybersecurity linguistic model trained from scratch using the Word2Vec (CBOW) algorithm.
Second, when using pretrained linguistic models (such as ELMo and BERT) without any fine-tuning, the NER model's results are inferior to the results of the supervised conditional random field algorithm, which does not utilize unlabeled data.          
These two observations support our main hypothesis that cybersecurity semantics are quite different than the semantics of the common English language.
We also note that there is a significant difference between the macro-average F-score of the NER model that is based on the cybersecurity linguistic model trained from scratch using the Word2Vec (CBOW) algorithm and the NER models that are based on fine-tuned  ELMo/BERT models.
This observation indicates that NER models based on fine-tuned linguistic models are less sensitive to the minority classes.
\end{description}

\begin{table*}
\begin{tabular}{|c|ccccccccc|cc|}
\Xhline{2\arrayrulewidth}
\multicolumn{1}{|c|}{}
&\multicolumn{1}{c}{\begin{tabular}[c]{@{}c@{}}Attack\\Vector\end{tabular}}
&\multicolumn{1}{c}{\begin{tabular}[c]{@{}c@{}}Attack\\Technique\end{tabular}}
&\multicolumn{1}{c}{\begin{tabular}[c]{@{}c@{}}Attack\\Impact\end{tabular}}
&\multicolumn{1}{c}{\begin{tabular}[c]{@{}c@{}}Attack\\Mean\end{tabular}}
&\multicolumn{1}{c}{\begin{tabular}[c]{@{}c@{}}Vulnerable\\Platform\end{tabular}}
&\multicolumn{1}{c}{\begin{tabular}[c]{@{}c@{}}Vulnerable\\OS\end{tabular}}
&\multicolumn{1}{c}{\begin{tabular}[c]{@{}c@{}}Vulnerable\\Version\end{tabular}}
&\multicolumn{1}{c}{\begin{tabular}[c]{@{}c@{}}Network\\Protocol\end{tabular}}
&\multicolumn{1}{c|}{\begin{tabular}[c]{@{}c@{}}Network\\Port\end{tabular}}
&\multicolumn{1}{c}{\begin{tabular}[c]{@{}c@{}}Micro \\  Average\end{tabular}}
&\multicolumn{1}{c|}{\begin{tabular}[c]{@{}c@{}}Macro \\ Average\end{tabular}}\\
\Xhline{2\arrayrulewidth}

\begin{tabular}[c]{@{}c@{}}CRF~\cite{lal2013information}\end{tabular}
 & 0.86 & 0.19 & 0.49 & 0.33 & 0.42 & 0.35 & 0.43 & 0.68 & 0.95 & 0.43 & 0.52 \\
\begin{tabular}[c]{@{}c@{}}CRF~\cite{weerawardhana2014automated}\end{tabular} 
& 0.93 & 0.60 & 0.78 & 0.55 & 0.72 & 0.66 & 0.81 & 0.66 & 0.94 & 0.77 & 0.73\\ 
\begin{tabular}[c]{@{}c@{}}Rule-based~\cite{weerawardhana2014automated}\end{tabular} 
& - & 0.50 & 0.59 & 0.33 & 0.47 & - & 0.49 & - & - & 0.47 & 0.46\\
\hline
\begin{tabular}[c]{@{}c@{}}ELMo (pretrained)\end{tabular} 
& 0.86 & 0.75 & 0.82 & 0.64 & 0.68 & 0.59 & 0.78 & 0.30 & 0.73 & 0.74 & 0.68 \\
\begin{tabular}[c]{@{}c@{}}BERT (pretrained)\end{tabular} 
& 0.79 & 0.61 & 0.73 & 0.58 & 0.62 & 0.42 & 0.68 & 0.30 & 0.35 & 0.63 & 0.56\\
\hline
\begin{tabular}[c]{@{}c@{}}ELMo (fine-tuned)\end{tabular} 
& 0.93 & 0.85 & 0.88 & 0.72 & 0.77 & 0.61 & 0.86 & 0.35 & 0.83 & 0.81 & 0.75 \\
\begin{tabular}[c]{@{}c@{}}BERT (fine-tuned)\end{tabular} 
& 0.92 & 0.78 & 0.86 & 0.72 & 0.73 & 0.66 & 0.78 & 0.59 & 0.40 & 0.80 & 0.71 \\
\Xhline{2\arrayrulewidth}
\begin{tabular}[c]{@{}c@{}}Word2Vec (Skip-Gram)\end{tabular}
& 0.94 & 0.87 & 0.89 & 0.74 & 0.76 & 0.72 & 0.72 & 0.43 & 0.80 & 0.80 & 0.76 \\
\begin{tabular}[c]{@{}c@{}}Word2Vec (CBOW)\end{tabular} 
& 0.94 & 0.89 & 0.90 & 0.78 & 0.76 & 0.75 & 0.80 & 0.65 & 0.94 & 0.83 & 0.82 \\

\Xhline{2\arrayrulewidth}

\begin{tabular}[c]{@{}c@{}}Word2Vec + CMI\\(CBOW + k-NN)\end{tabular} 
& \textcolor{red}{0.95} & 0.89 & \textcolor{red}{0.92} & \textcolor{red}{0.80} & 0.76 & 0.75 & 0.80 & 0.65 & 0.94 & 
\textcolor{red}{0.83} & \textcolor{red}{0.82} \\

\begin{tabular}[c]{@{}c@{}}Word2Vec + CMI\\(CBOW + LR)\end{tabular} 
& \textcolor{red}{0.96} & 0.89 & \textcolor{red}{0.93} & \textcolor{red}{0.82} & 0.76 & 0.75 & 0.80 & 0.65 & 0.94 & 
\textcolor{red}{0.84} & \textcolor{red}{0.83} \\
\Xhline{2\arrayrulewidth}

\end{tabular}
\caption{Performance comparison (F1-scores) of the different models in attack entity extraction task.\label{table-ner-results}}
\end{table*}

\section{\label{sec:complete_missing_info}Complete Missing Information}
The output of the attack entity extraction algorithm occasionally misses valuable information about the attack; this is due to the following reasons:
In some cases, the vulnerability description itself is incomplete.
For example, the description presented in Figure~\ref{fig:cve-example} does not include information regarding the context required by the attacker for successful exploitation of the vulnerability (i.e., attack vector).
In other cases, although the vulnerability description includes all of the information about the attack, the attack entity extraction model fails to identify the attack entity.
In both cases, this information, which is relevant for modeling the attack, can often be completed (predicted) based on similar, previously labeled, vulnerabilities.
 
Formally, we define the problem of completing the missing information as follows:
Let $c_i$ be a security vulnerability description, $E_{c_i}^*$ be the ground truth labels (entities) of $c_i$, and $E_{c_i} \subset E_{c_i}^*$ be the subset of entities that are identified by the attack entity extraction algorithm.
The task of completing the missing information of $c_i$ is predicting $\hat{E_{c_i}} = E_{c_i}^* \setminus E_{c_i}$.

We propose an algorithm for predicting the attack entities that are missing.
The proposed method utilizes the set of entities that are identified by the attack entity extraction algorithm $E_{c_i}$ to predict $\hat{E_{c_i}}$.
We specifically focus on the following three attack entities: attack vector, attack impact, and attack means, which are important for modeling an attack.

\subsection{\label{sec:complete_missing_info-representation}Dataset}
The dataset is based on the unlabeled dataset used to create the cybersecurity linguistic model.
As presented in Section \ref{sec:embedding}, this dataset consists of 146K vulnerability descriptions from the NVD.
We applied the attack entity extraction model (when using the Word2Vec-CBOW cybersecurity linguistic model) on the 146K vulnerability descriptions from the NVD and filter vulnerabilities with less than four attack entities and vulnerabilities that do not include the three attack entities described above (i.e., attack vector, attack impact, and attack means).
The filtered dataset includes 40K vulnerability descriptions; we randomly selected 30K vulnerability descriptions as the training set, and the reminder served as the test set.

\subsection{\label{sec:complete_missing_info-algo}Complete missing information algorithm}
\begin{description}[style=unboxed,leftmargin=0cm]
\item[\textbf{Feature representation}.]
The process of creating the feature vector for a vulnerability description includes the following three phases:
\begin{itemize}
    \item \textit{Extract.}
    The feature vector used for predicting the missing information is based on the set of known entities (denoted by $E_{c_i}$).
    Specifically, we apply the attack entity extraction model (when using the Word2Vec-CBOW cybersecurity linguistic model) to extract the attack entities from the vulnerability description.
    The output of this phase is a list of attack entities and their corresponding values (words), which are written in free text.
    
    \item \textit{Transform.}
    In this phase, we transform the free text values of each attack entity into numerical vectors.
    This is done by using the Word2Vec-CBOW cybersecurity linguistic model.
    The output of this phase is the embedding representation for every value (word) in any attack entity.
    Please note that the number of values in each attack entity is not the same for each vulnerability.
    Therefore, the values of each attack entity are represented using a different number of vectors (each of which is the size of the embedding dimension $D$).
    
    \item \textit{Aggregate.}
    In this phase, we aggregate the vectors that represent the values of each attack entity into a single vector.
    We refer to this vector as the succinct vector (SV).
    The SV is calculated based on the following equation:
    \begin{equation}\label{eq:entity_vector}
    SV(v_1,\ldots,v_k) = \frac {\sum_{i=1}^{k} \frac{v_i}  {\small\Vert v_i \small\Vert}} {k} 
    \end{equation}
    The output of this phase consists of the SVs for each attack entity.
\end{itemize}

    The concatenation of the SVs is the feature vector used by the complete missing information algorithm.
    The feature vector includes 900 features (i.e., 100 features for each attack entity).
    In a case in which the attack entity has no value, the corresponding features are equal to zero.\\
    
\item[\textbf{Algorithm}.]
The proposed approach for completing the missing information is based on $k$-means clustering for class discretization and logistic regression for prediction.
\begin{itemize}
    \item \textit{Class Discretization.}
    The task of completing the missing information for a security vulnerability is essentially predicting the specific values of the missing attack entities. 
    Since attack entities are written in free text, the same sentence can be written in various ways.
    For instance, the following phrases: `execute arbitrary code,' `code execution,' `code exec,' and `running code' have the same meaning and therefore should be treated similarly. 
    However, their representations (using the SV) are slightly different.
    In addition, training an algorithm on a 100-dimensional output space is difficult.
    To mitigate this problem, we use the k-means clustering algorithm to cluster the output values into groups having the same meaning.
    The results of this clustering for the tree attack entity (namely, attack vector, attack impact, and attack means) are presented in Figure~\ref{fig:entities_clusters}.
    As can be seen, each attack entity has very clear clusters. 
    We labeled the clusters manually by analyzing several samples from each cluster.
    This labeling was performed based on MulVAL's existing primitives. 
    
\begin{figure}[t!]
     \centering
     \begin{subfigure}[t]{0.4\textwidth}
         \centering
         \includegraphics[width=\textwidth]{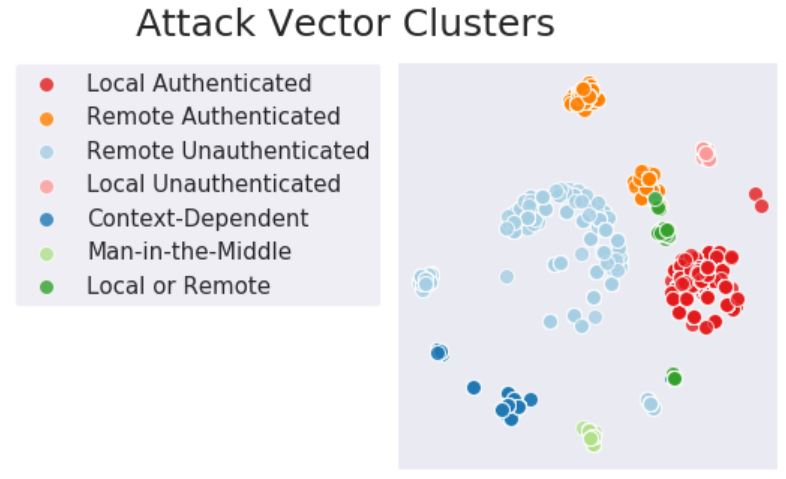}
         \label{fig:sub_av-clusters}
     \end{subfigure}
     \hfill
     \begin{subfigure}[t]{0.4\textwidth}
         \centering
         \includegraphics[width=\textwidth]{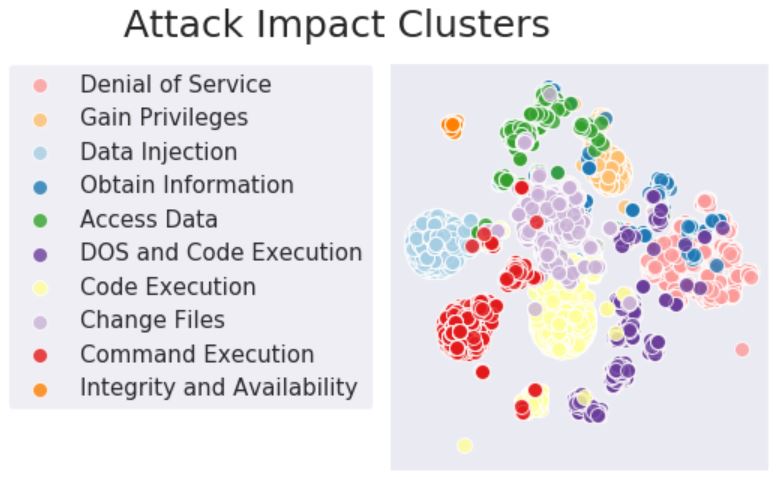}
         \label{fig:sub_impact}
     \end{subfigure}
     \hfill
          \begin{subfigure}[t]{0.4\textwidth}
         \centering
         \includegraphics[width=\textwidth]{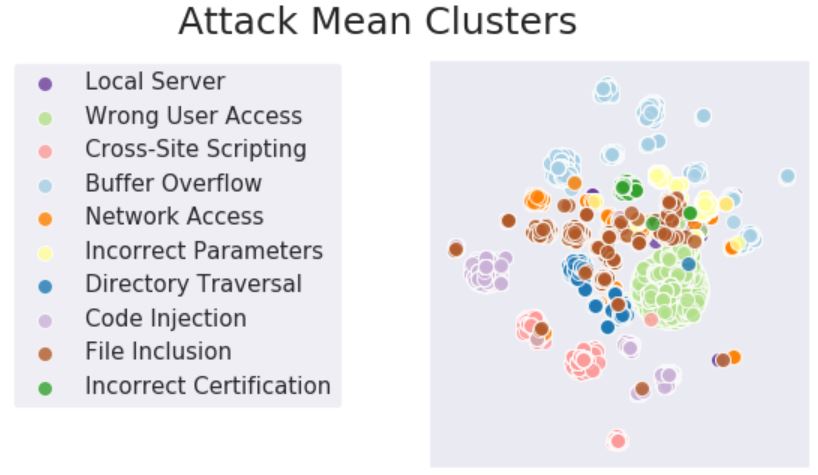}
         \label{fig:sub_mean}
     \end{subfigure}
     \hfill

        \caption{The clusters of each entity that we predict.}
        \label{fig:entities_clusters}
\end{figure}

    \item \textit{Prediction.}
    The prediction model is based on the logistic regression algorithm.
    This algorithm measures the relationship between the categorical dependent variable (i.e., the cluster number) and the independent variables (feature vector) by estimating coefficients $w_k$ using the following multinominal logistic function.
    
\begin{equation}\label{eq:regression}
    f(X) = \operatorname*{argmax}_k \frac{e^{w_{k}X+b_k}}{\sum_{j=1}^{K} e^{w_{k}X+b_k}} 
\end{equation}
where $k$ is the cluster number, $X$ is the feature vector, and $w_k,b_k$ are the model coefficients and bias respectively. \\
\end{itemize}

\item[\textbf{Optimization.}]
We trained the model using the limited memory BFGS optimization algorithm and cross-entropy loss for 70 iterations. 
During the training we use $l2$ regularization.

\end{description}

\subsection{Evaluation}
\begin{description}[style=unboxed,leftmargin=0cm]
\item[Experimental setup.]
We tested the proposed algorithm and compared it to a simple k-nearest neighbors (k-NN) algorithm without class discretization, which served as the baseline.
In the k-NN model, the instance was classified based on the values of the attack entity of its nearest neighbor.
We trained three logistic regression models and three k-NN models to predict the missing values for each of the three attack entities. 
The models were trained on the dataset described above, which includes 30K vulnerability descriptions from the NVD and tested on the remaining 10K vulnerability descriptions.
In order to train the model, we simulated the existence of missing entities by removing the SVs that represent the label (from the feature vector).
This trick, known as masking, is very common when labeled data is not available.
All of the experiments were performed on a 64-bit Windows Server 2008 R2 Enterprise machine, with a 2.00 GHZ Intel Xeon CPU (version E5-2620 with 24 logical cores) and 64 GB of RAM.
The algorithms were implemented in Python using the sklearn library. 

\item[Evaluation measures.] 
In order to evaluate the model, we defined the following three evaluation measures:
\begin{itemize}
    \item \textit{Performance on masked attack entities - } evaluates the performance of the two algorithms in predicting the masked values.
    We used the recall and precision at $k$ as measures for the accuracy of the models,
    where $k$ denotes the number of values recommended by the model.
    These metrics are very common in the field of recommendation systems.
    
    \item \textit{Performance on attack entity extraction -} evaluates the performance of the two algorithms in improving the attack entity extraction algorithm.
    Specifically, we tested whether the two algorithms can reduce the number of false negatives of the attack entity extraction algorithm.
    Similar to Section~\ref{sec:attack-entity-recog}, we used the F-score as a measure for the accuracy.
\end{itemize}

\item[Results.]
In this section, we present the results with respect to the two evaluation measures. 
\begin{itemize}
    \item \textit{Performance on masked attack entities.}
    The results, which are presented in Figure \ref{fig:final_res}, are very promising.
    Specifically, the logistic regression model outperforms the k-NN model for all three attack entities.
    
    \begin{figure}[t!]
    \centering
    \includegraphics[width=0.5\textwidth]{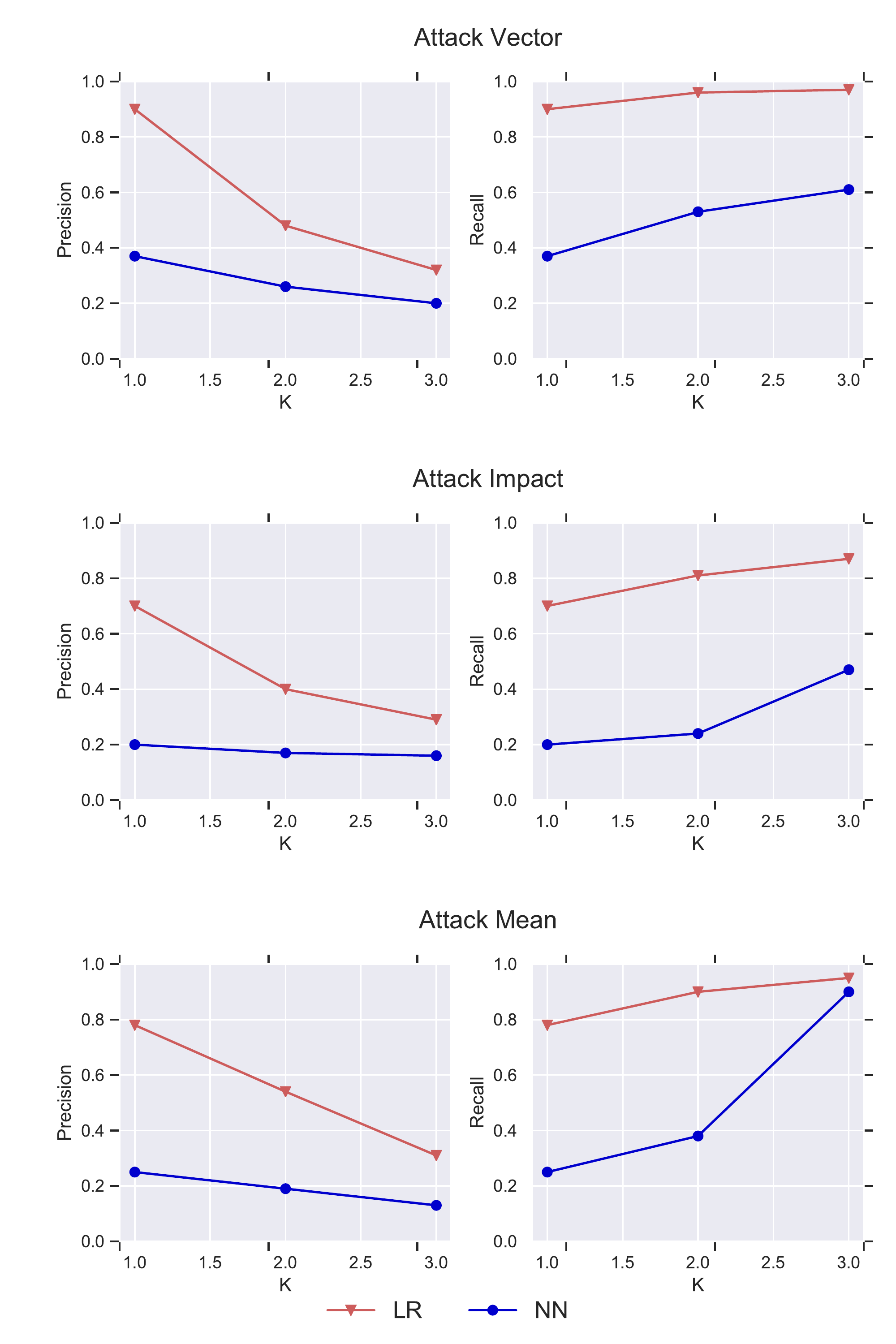}
    \caption{Performance comparison (recall and precision at $k$) of the two algorithms in the task of completing the missing information (evaluation on masked attack entities).}
    \label{fig:final_res}
\end{figure}

    \item \textit{Performance on attack entity extraction.}
    The results are presented in Figure~\ref{fig:final_res}.
    As can be seen, the process of completing the missing information improves the performance of the Word2Vec (CBOW) model in attack entity extraction.
    Specifically, when using the logistic regression model for completing the missing information, the average performance improves by 1\%.

\end{itemize}
\end{description}

\section{\label{sec:app}Generate \mulval Interaction Rules}
In this section, we present the process of automatically generating \mulval's interaction rules, given the entities extracted for a security vulnerability (phase 4 in Figure~\ref{fig:cve-example}).
We begin by providing an introduction to the \mulval framework.
Then, we present the process of automatically generating the interaction rules. 
Finally, we present an end-to-end evaluation for the entire process of generating \mulval's interaction rules, given a description of a security vulnerability.

\subsection{Introduction to logical attack graphs and \mulval}

\begin{description}[style=unboxed,leftmargin=0cm]
\item[Logical attack graphs.] A logical attack graph is defined as a tuple $AG = (N_r, N_p, N_d, E, \mathcal{L}, \mathcal{G})$, where:
\begin{itemize}
    \item $N_r$ is the set of derivation nodes that correspond to rules and represent the logic for a fact to become true (visualized as circles); these nodes imply an \textit{AND} relation between their incoming nodes;
    \item $N_p$ is the set of primitive facts that correspond to the inputs describing the specific system (visualized as rectangles);
    \item $N_d$ is the set of derived facts that are the results of applying rules on the primitive facts (visualized as diamonds); these nodes imply an \textit{OR} relation between their incoming nodes;
    \item $E \subseteq \{(N_p \cup N_d)\times N_r\} \cup \{N_r \times N_d\}$ is the set of edges (i.e., an edge can connect a primitive or derived fact with a derivation node or a derivation node with a derived fact);
    \item $\mathcal{L}$ is a mapping between nodes and their labels; and 
    \item $\mathcal{G}$ is the node that represents the attacker's goal.
\end{itemize}

Note that the incoming nodes to a derivation node represent the preconditions for performing the corresponding actions, while the incoming nodes to a derived node represent the various paths that lead to the same consequence.

\item[The MulVAL framework.]
\mulval~\cite{ou2005mulval} is a logic-based network security analyzer, which provides automated attack graph generation and analysis.
The modeling language used by MulVAL is Datalog~\cite{ou2005mulval}, which is a subset of the Prolog logic programming language.
The Datalog language consists of \emph{facts} and \emph{rules}, which are defined using \emph{predicates}. 

A predicate is an atomic formula of the form: $p(t_1,...,t_k)$, where each argument $t_i$ can be either a constant (starting with a lowercase letter) or a variable (starting with an uppercase letter). 
For instance, the following predicate states that some vulnerability is present in the \textit{oracleDB} program running on \textit{dbServer}: $vulExists(dbServer, VulID, oracleDB).$

Rules (referred to as \textit{interaction rules} in MulVAL) are represented using Horn clauses as follows: $P_0 :- P_1,...,P_n$, which essentially indicates that if the predicates $P_1,...,P_n$ are true, then predicate $P_0$ is also true. 
The left part of the clause ($P_0$) is referred to as the \textit{head}, and the right part ($P_1,...,P_n$) is referred as the \textit{body}.

Facts are clauses without a body; they are used to specify attacker and system states.
For instance, the attacker’s goal and initial location, host and network configuration, and existing vulnerabilities.
While facts can be derived automatically using tools, such as Nmap, Nessus, and OpenVAS, designing and creating new interaction rules is not a challenging task which is performed manually using security experts.

In order to execute a Datalog program, MulVAL uses the XSB environment, which is an extended implementation of the Prolog programming language that supports \emph{tabled execution}. 
Tabled execution prevents the recalculation of previously calculated facts (i.e., each fact is calculated only once). 
Since the number of facts is polynomial in the size of the network, executing a Datalog program has polynomial time complexity.

\end{description}

\subsection{\label{sec:app-rules}Algorithm for automatic rule generation}
The process of automatically generating \mulval's interaction rules consists of three main phases (phase 4 in Figure~\ref{fig:cve-example}): structure creation (phase 4a), constant argument assignment (phase 4b), and variable wiring (phase 4c).

\begin{description}[style=unboxed,leftmargin=0cm]
\item[Structure creation.]
In this phase, we create the structure of the interaction rule, given the entities and values extracted for the security vulnerability.
As mentioned, the structure of an interaction rule includes a head and a body.
The head represents the impact of the attack, and the body represents the list of preconditions required by an attacker to achieve this impact.

In order to derive the structure of the interaction rule, we utilize three attack entities: attack impact, attack mean, and attack vector.
Based on the output of the attack entity extraction algorithm (Section~\ref{sec:attack-entity-recog}), we identify (within the vulnerability description) the entities that specify the impact and mean of the attack, as well as the context in which attack exploitation is possible (i.e., the attack vector).
In a case in which some of the entities are missing, we apply the logistic regression model for completing missing information (Section~\ref{sec:complete_missing_info}).
Note that if this information cannot be completed by the algorithm, the rule cannot be created.  

After determining the text that is relevant to each of those attack entities, we apply the relevant k-means clustering model to map the free text values into predicates.
The predicate that specifies the impact of the attack defines the head of the interaction rule, and the predicates that specify the attack vector and attack mean specify the body of the interaction rule (see phase 4a in Figure \ref{fig:cve-example}).

\item[Constant argument assignment.]
As mentioned, interaction rules support two types of arguments: constant and variable arguments. 
After defining the structure of the interaction rule, we assign values to the constant arguments.
This is done based on the extracted attack entities (see phase 4b in Figure \ref{fig:cve-example}).
We map each constant argument to the relevant attack entities and use the information extracted from the vulnerability description to set this argument.

\item[Variable wiring.]
Predicates may include large sets of variables.
In contrast to constant arguments, which are bound when formulating the predicate, the binding of a variable is performed during the pattern matching process used by the Datalog interpreter.
However, when formulating a new interaction rule, there is a need to correctly wire the variables across the different predicates of the interaction rule.
Wiring two variables is performed by assigning them the same name.

For instance, in the example presented in Figure~\ref{fig:cve-example} (phase 4c), we can identify multiple variables that specify a \textit{host} (e.g., Principal, Host, SrcHost, DstHost).
However, in some predicates this host represents the vulnerable host (such as in the networkService predicate), and in other predicates this host represents the attacker host (such as in the attackerLocated predicate).
Furthermore, there are predicates for which both the attacker host and target host should be specified (such as in execCode and netAcsess).

It should be mentioned that wiring the variable correctly is crucial for producing an interaction rule that is 
semantically correct.
In this phase, we utilize an algorithm that can be used to automatically wire the variables of new interaction rules.
We use a probability matrix that represents the probability for two variables to be wired.
This probability is calculated based on the wiring observed in existing interaction rules.

Formally, the probability matrix $M\in \mathbb{R}^{n\times n}$ is a symmetric matrix that represents all of the variables that exist in the initial rule set.
The value $M[i,j]$ represents the probability for wiring the $i^th$ and $j^th$ variables, given that they exist in the same interaction rule.
This probability can be estimated by dividing the number of times these two variables were wired by the number of times these two variable appeared together in the same interaction rule. 

Using the probability matrix, we can estimate the probability for wiring two variables.
However, since many variables never appear together in the same interaction rule, this matrix includes unknown values.
In order to solve this problem, we used a k-nearest neighbors (k-NN) imputer model.
Specifically, when facing an unknown value, we complete that value with the average across the K closest vectors in the matrix.
\end{description}

\subsection{Evaluation}
\begin{figure}[ht]
\begin{subfigure}{.5\textwidth}
  \centering
  \includegraphics[width=0.9\linewidth]{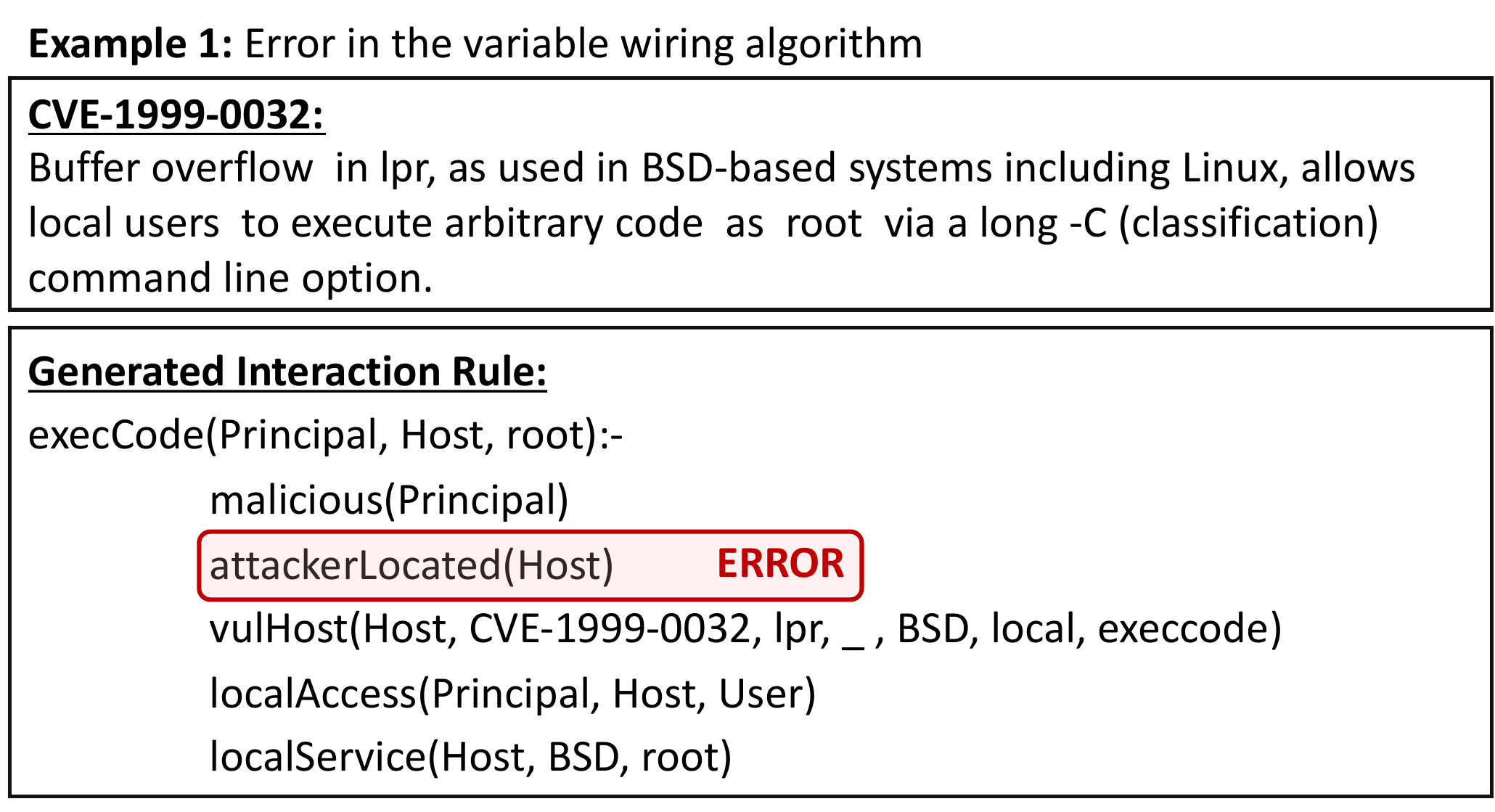}  
  \caption{The \textit{attackerLocated} predicate specifies the host of the attacker, which represented using the \textit{Principal} variable.
}
  \label{fig:error1}
\end{subfigure}
\begin{subfigure}{.5\textwidth}
  \centering
  \includegraphics[width=0.9\linewidth]{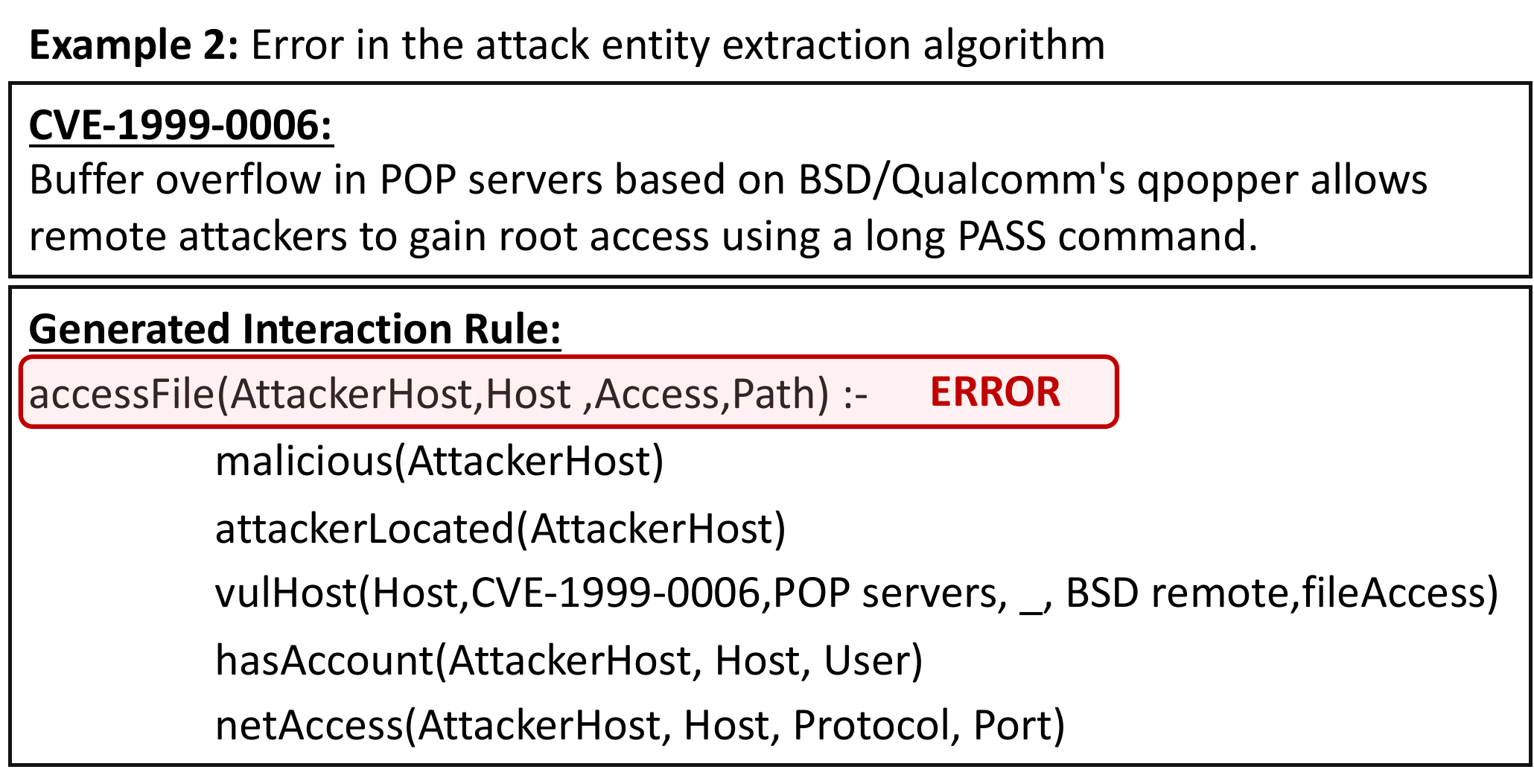}  
  \caption{The attack entity extraction algorithm misclassified the impact of that attack as \textit{access} instead of \textit{gain root access}. Following that, the class discretization algorithm mapped this value to \textit{Access Data}}.
  \label{fig:sub-second}
\end{subfigure}
\caption{Examples of errors made by the automatic generation of interaction rules.}
\label{fig:error_analyis}
\end{figure}

\begin{description}[style=unboxed,leftmargin=0cm]
\item[Experimental setup.]
We conducted two different experiments.
The first experiment focused on evaluating the variable wiring model. 
The dataset used for this experiment is based on the extended attack graph modeling presented in~\cite{stan2019extending}.
This dataset includes 199 interaction rules.
The model was trained/tested using a 10-fold cross-validation setup.

The second experiment focused on evaluating the entire framework (end-to-end evaluation).
The dataset used for this experiment includes 27K randomly selected vulnerabilities from the NVD.

All of the experiments were performed on a 64-bit Windows Server 2008 R2 Enterprise machine, with a 2.00 GHZ Intel Xeon CPU (version E5-2620 with 24 logical cores) and 64 GB of RAM.
The  variable wiring algorithm was implemented in Python using the sklearn library.

\item[Evaluation measures.]
In the first experiment, we used the average F-score~\cite{tharwat2018classification} as a measure for the accuracy of the models.

In the second experiment, we used the \textit{success ratio} as a measure for the accuracy of the model.
We define the success ratio as the portion of vulnerability descriptions for which our proposed framework was able to derive interaction rules.

In addition to the \textit{success ratio}, we manually validated 50 randomly selected rules. 

\item[Results.]
The results in the first experiment -- evaluation of the variable wiring model -- are promising, with an average F1-score of 0.84 (and an average accuracy of 0.97).

In the end-to-end experiment, the framework was able to automatically generate interaction rules for 19.5K of the 27K vulnerabilities that were processed (i.e., a success ratio of 72\%).
When analyzing the vulnerabilities that could not be modeled by the framework, we observed that the main reason for the framework's inability to generate a rule is missing values for the attack mean and attack impact entities.
In addition, for about 1,300 of the vulnerabilities that could not be processed, the attack mean was described as an `unspecified vulnerability.'

Our manual assessment of the 50 randomly selected rules indicates that 36 out of the 50 rules were created correctly (72\%).
The main errors identified in the incorrectly generated rules are: missclassification of the \textit{impact} or \textit{mean} of the attack by the attack entity extraction algorithm (seven rules); incorrectly wiring the variables of the predicates by the variable wiring algorithm (three rules); and a missing impact that was completed incorrectly by the completing missing information algorithm (four rules).
Examples of these errors are presented in Figure~\ref{fig:error_analyis}.

\end{description}

\section{\label{sec:related_work}Related Work}

Previous studies related to our research have focused mainly on the first challenge -- extracting cybersecurity information from text.

In the domain of entity extraction, previous works utilized different data sources; the most commonly used is the NVD~\cite{aksu2018automated, weerawardhana2014automated, mulwad2011extracting, bridges2013automatic, gasmi2018lstm, lal2013information}.
The national vulnerability database (NVD) is a database that contains common vulnerabilities and exposures (CVEs). 
The information provided for each vulnerability within the NVD includes its description, references (list of URLs), date of entry creation, etc.\\
Another commonly used dataset is the Metasploit database~\cite{lal2013information, bridges2013automatic, gasmi2018lstm}. 
The Metasploit project is an open-source project that serves as a public resource for researching security vulnerabilities and developing code.
For each exploit the Metasploit database includes the description, CVE-ID (if it exists), score, references (list of URLs), etc.\\
Other common data sources are security blogs~\cite{lal2013information, weerawardhana2014automated, more2012knowledge}, antivirus websites~\cite{lal2013information}, APT reports~\cite{li2019self}, CTI reports~\cite{kim2020automatic}, and vulnerability datasets of specific companies, such as Microsoft~\cite{weerawardhana2014automated,bridges2013automatic,gasmi2018lstm} and Adobe~\cite{weerawardhana2014automated}.

Various methods have being proposed for extracting cybersecurity-related information from semi-structured/unstructured datasets.
The most simple and naive method is a rule-based approach~\cite{jing2017augmenting, weerawardhana2014automated, aksu2018automated}.
In this approach, all of the rules are manually defined, and therefore this approach cannot scale.
In the proposed framework, we utilize an unsupervised machine learning approach, and therefore our framework does not rely on the manual efforts of security experts.

Classic machine learning models (e.g., CRF, SVM, and HMM) were also used for identifying attack entities' values~\cite{aksu2018automated, bridges2013automatic, weerawardhana2014automated, jones2015towards}.
These models rely on handcrafted (e.g., orthographic) features that are defined manually by security experts.
In addition, these methods can only process a single data point (i.e., word) and are unable to capture the entire sequence (i.e., a sentence); therefore, they cannot be used to capture long-term dependencies.

With the emerging use of deep neural networks (DNNs) in the NLP domain, analyzing cybersecurity-related text using NLP techniques became popular~\cite{li2019self, simran2019deep, kim2020automatic, gasmi2019cold}.
Previous research used word embeddings~\cite{simran2019deep, gasmi2019cold, li2019self} or character embeddings~\cite{li2019self} as the feature for representing entities.
These works utilized pretrained embedding methods that were trained on general English datasets.
Since the cybersecurity domain includes specific terminology and linguistic semantics which are different from general English repositories, these methods are less accurate than our proposed method which is trained on the NVD dataset.
Mulwad \textit{et al.}~\cite{mulwad2011extracting} used OpenCalais and Wikitology, which are general (not cybersecurity-related) tools for named entity recognition.
An analysis of the results obtained by these tools when they are applied on cybersecurity-related texts with a set of manually defined rules, shows that they are able to identify the security-related entities within the text.
While this method utilized existing NLP tools, it still requires a human expert to define the rules for post-processing.

Unlike general NLP tasks, in the domain of cybersecurity,
there is no predefined set of entities to predict that is commonly used in all studies in this area.
For example, while the authors of~\cite{lal2013information} and \cite{weerawardhana2014automated} extracted six types of entities that describe attacks, the authors of~\cite{aksu2018automated} focused on extracting the privilege entity of attacks (i.e., the privileges required for the attack to be successful), Bridges \textit{et al.}~\cite{bridges2013automatic}
focused on entities that are related to the the attacker action, and Mulwad \textit{et al.}~\cite{mulwad2011extracting} focused on the vulnerability type.
In this work, since we are aiming to derive the interaction rules that model the attacks, we extract a comprehensive set of 11 different types of entities.

We summarize and compare the prior work mentioned above in Table~\ref{table-works} in the Appendix.

While almost all of the studies presenting methods to extract cybersecurity entities from textual sources motivated their research with the need to model attacks~\cite{jing2017augmenting, aksu2018automated}, only two of them presented methods for achieving this goal \cite{gasmi2019cold, mulwad2011extracting}.
In~\cite{jing2017augmenting}, the authors derived a model for generating \mulval interaction rules, however this model has two main limitations.
First the CVE's vulnerability type is known prior to the entity extraction.
In our work, the model doesn't need to know the vulnerability type in advance in order to generate the right rule.
Second, the authors did not mention how to wire the variables across the different predicates.

To the best of our knowledge, our work is the first to present an end-to-end fully automated process for deriving interaction rules from textual descriptions of security vulnerabilities, as well as the first to propose methods for completing missing attack information.

\section{\label{sec:conclusion}Conclusions and Future Work}

We present a novel, end-to-end, automated framework for modeling new attack techniques from textual description of a security vulnerability.
The proposed framework implements a novel pipeline that includes a dedicated cybersecurity linguistic model trained on the the NVD repository, a recurrent neural network model used for attack entity extraction, a  logistic regression model used for completing the missing information, and a novel machine learning-based approach for automatically modeling the attacks as \mulval's interaction rule.
We evaluated the performance of each of the individual algorithms, as well as of the complete framework and demonstrated its effectiveness.

In future work we plan to evaluate the impact of the new attack interaction rules on the attack graphs derived for different organization networks (both in terms of run time and new attack paths). 
In addition, we plan to extend our cybersecurity linguistic model to other cybersecurity data sources such as security blogs, CTI reports, etc. 
Finally, since some of the rules were created with errors, we plan to develop and provide a confidence score for each rule created that will indicate the level of assurance in the correctness of the rule.

\clearpage
\bibliographystyle{IEEEtranS}
\bibliography{main}

% Generated by IEEEtranS.bst, version: 1.12 (2007/01/11)
\begin{thebibliography}{10}
\providecommand{\url}[1]{#1}
\csname url@samestyle\endcsname
\providecommand{\newblock}{\relax}
\providecommand{\bibinfo}[2]{#2}
\providecommand{\BIBentrySTDinterwordspacing}{\spaceskip=0pt\relax}
\providecommand{\BIBentryALTinterwordstretchfactor}{4}
\providecommand{\BIBentryALTinterwordspacing}{\spaceskip=\fontdimen2\font plus
\BIBentryALTinterwordstretchfactor\fontdimen3\font minus
  \fontdimen4\font\relax}
\providecommand{\BIBforeignlanguage}[2]{{%
\expandafter\ifx\csname l@#1\endcsname\relax
\typeout{** WARNING: IEEEtranS.bst: No hyphenation pattern has been}%
\typeout{** loaded for the language `#1'. Using the pattern for}%
\typeout{** the default language instead.}%
\else
\language=\csname l@#1\endcsname
\fi
#2}}
\providecommand{\BIBdecl}{\relax}
\BIBdecl

\bibitem{Nessuscite}
``R. deraison, nessus, retrieved may 2003, from http://www.nessus.org, 1999.''
  1999.

\bibitem{aksu2018automated}
M.~U. Aksu, K.~Bicakci, M.~H. Dilek, A.~M. Ozbayoglu \emph{et~al.}, ``Automated
  generation of attack graphs using nvd,'' in \emph{Proceedings of the Eighth
  ACM Conference on Data and Application Security and Privacy}.\hskip 1em plus
  0.5em minus 0.4em\relax ACM, 2018, pp. 135--142.

\bibitem{albanese2014manipulating}
M.~Albanese, E.~Battista, S.~Jajodia, and V.~Casola, ``Manipulating the
  attacker's view of a system's attack surface,'' in \emph{2014 IEEE Conference
  on Communications and Network Security}.\hskip 1em plus 0.5em minus
  0.4em\relax IEEE, 2014, pp. 472--480.

\bibitem{bakarov2018survey}
A.~Bakarov, ``A survey of word embeddings evaluation methods,'' \emph{arXiv
  preprint arXiv:1801.09536}, 2018.

\bibitem{bengio1994learning}
Y.~Bengio, P.~Simard, P.~Frasconi \emph{et~al.}, ``Learning long-term
  dependencies with gradient descent is difficult,'' \emph{IEEE transactions on
  neural networks}, vol.~5, no.~2, pp. 157--166, 1994.

\bibitem{bikel1999algorithm}
D.~M. Bikel, R.~Schwartz, and R.~M. Weischedel, ``An algorithm that learns
  what's in a name,'' \emph{Machine learning}, vol.~34, no. 1-3, pp. 211--231,
  1999.

\bibitem{borthwick1999maximum}
A.~Borthwick and R.~Grishman, ``A maximum entropy approach to named entity
  recognition,'' Ph.D. dissertation, Citeseer, 1999.

\bibitem{bridges2013automatic}
R.~A. Bridges, C.~L. Jones, M.~D. Iannacone, K.~M. Testa, and J.~R. Goodall,
  ``Automatic labeling for entity extraction in cyber security,'' \emph{arXiv
  preprint arXiv:1308.4941}, 2013.

\bibitem{curran2003language}
J.~R. Curran and S.~Clark, ``Language independent ner using a maximum entropy
  tagger,'' in \emph{Proceedings of the seventh conference on Natural language
  learning at HLT-NAACL 2003}, 2003, pp. 164--167.

\bibitem{developers2012open}
O.~Developers, ``The open vulnerability assessment system (openvas),'' 2012.

\bibitem{devlin2018bert}
J.~Devlin, M.-W. Chang, K.~Lee, and K.~Toutanova, ``Bert: Pre-training of deep
  bidirectional transformers for language understanding,'' \emph{arXiv preprint
  arXiv:1810.04805}, 2018.

\bibitem{elsahar_2017}
\BIBentryALTinterwordspacing
H.~Elsahar, ``T-rex : A large scale alignment of natural language with
  knowledge base triples [json sample],'' Jun 2017. [Online]. Available:
  \url{https://figshare.com/articles/dataset/T-Rex_A_Large_Scale_Alignment_of_Natural_Language_with_Knowledge_Base_Triples_JSON_SAMPLE_/5151175/1}
\BIBentrySTDinterwordspacing

\bibitem{fila2020exploiting}
B.~Fila and W.~Wide{\l}, ``Exploiting attack–defense trees to find an optimal
  set of countermeasures,'' in \emph{The Proceedings of the 33rd IEEE Computer
  Security Foundations Symposium (CSF'20)}.\hskip 1em plus 0.5em minus
  0.4em\relax IEEE, 2020.

\bibitem{gasmi2018lstm}
H.~Gasmi, A.~Bouras, and J.~Laval, ``Lstm recurrent neural networks for
  cybersecurity named entity recognition,'' \emph{ICSEA 2018}, p.~11, 2018.

\bibitem{gasmi2019cold}
H.~Gasmi, J.~Laval, and A.~Bouras, ``Cold-start cybersecurity ontology
  population using information extraction with lstm,'' in \emph{2019
  International Conference on Cyber Security for Emerging Technologies
  (CSET)}.\hskip 1em plus 0.5em minus 0.4em\relax IEEE, 2019, pp. 1--6.

\bibitem{graves2013speech}
A.~Graves, A.-r. Mohamed, and G.~Hinton, ``Speech recognition with deep
  recurrent neural networks,'' in \emph{2013 IEEE international conference on
  acoustics, speech and signal processing}.\hskip 1em plus 0.5em minus
  0.4em\relax IEEE, 2013, pp. 6645--6649.

\bibitem{inokuchi2019design}
M.~Inokuchi, Y.~Ohta, S.~Kinoshita, T.~Yagyu, O.~Stan, R.~Bitton, Y.~Elovici,
  and A.~Shabtai, ``Design procedure of knowledge base for practical attack
  graph generation,'' in \emph{Proceedings of the 2019 ACM Asia Conference on
  Computer and Communications Security}, 2019, pp. 594--601.

\bibitem{jing2017augmenting}
J.~T.~W. Jing, L.~W. Yong, D.~M. Divakaran, and V.~L. Thing, ``Augmenting
  mulval with automated extraction of vulnerabilities descriptions,'' in
  \emph{TENCON 2017-2017 IEEE Region 10 Conference}.\hskip 1em plus 0.5em minus
  0.4em\relax IEEE, 2017, pp. 476--481.

\bibitem{jones2015towards}
C.~L. Jones, R.~A. Bridges, K.~M. Huffer, and J.~R. Goodall, ``Towards a
  relation extraction framework for cyber-security concepts,'' in
  \emph{Proceedings of the 10th Annual Cyber and Information Security Research
  Conference}, 2015, pp. 1--4.

\bibitem{kim2020automatic}
G.~Kim, C.~Lee, J.~Jo, and H.~Lim, ``Automatic extraction of named entities of
  cyber threats using a deep bi-lstm-crf network,'' \emph{INTERNATIONAL JOURNAL
  OF MACHINE LEARNING AND CYBERNETICS}, 2020.

\bibitem{kordy2017how}
B.~Kordy and W.~Wide{\l}, ``How well can i secure my system?'' in
  \emph{Integrated Formal Methods}, N.~Polikarpova and S.~Schneider, Eds.\hskip
  1em plus 0.5em minus 0.4em\relax Cham: Springer International Publishing,
  2017, pp. 332--347.

\bibitem{kordy2018on}
B.~Kordy and W.~Wide\l, ``On quantitative analysis of attack--defense trees
  with repeated labels,'' in \emph{Principles of Security and Trust}, L.~Bauer
  and R.~K{\"u}sters, Eds.\hskip 1em plus 0.5em minus 0.4em\relax Cham:
  Springer International Publishing, 2018, pp. 325--346.

\bibitem{lal2013information}
R.~Lal \emph{et~al.}, ``Information extraction of security related entities and
  concepts from unstructured text.'' 2013.

\bibitem{landoll2005security}
D.~J. Landoll and D.~Landoll, \emph{The security risk assessment handbook: A
  complete guide for performing security risk assessments}.\hskip 1em plus
  0.5em minus 0.4em\relax CRC Press, 2005.

\bibitem{li2019self}
T.~Li, Y.~Guo, and A.~Ju, ``A self-attention-based approach for named entity
  recognition in cybersecurity,'' in \emph{2019 15th International Conference
  on Computational Intelligence and Security (CIS)}.\hskip 1em plus 0.5em minus
  0.4em\relax IEEE, 2019, pp. 147--150.

\bibitem{nmap}
G.~Lyon, ``Nmap network scanning,'' 1997.

\bibitem{mann1999towards}
D.~E. Mann and S.~M. Christey, ``Towards a common enumeration of
  vulnerabilities,'' in \emph{2nd Workshop on Research with Security
  Vulnerability Databases, Purdue University, West Lafayette, Indiana}, 1999.

\bibitem{mccallum2003early}
A.~McCallum and W.~Li, ``Early results for named entity recognition with
  conditional random fields, feature induction and web-enhanced lexicons,''
  2003.

\bibitem{mcnamee2002entity}
P.~McNamee and J.~Mayfield, ``Entity extraction without language-specific
  resources,'' in \emph{COLING-02: The 6th Conference on Natural Language
  Learning 2002 (CoNLL-2002)}, 2002.

\bibitem{mikolov2013efficient}
T.~Mikolov, K.~Chen, G.~Corrado, and J.~Dean, ``Efficient estimation of word
  representations in vector space,'' \emph{arXiv preprint arXiv:1301.3781},
  2013.

\bibitem{more2012knowledge}
S.~More, M.~Matthews, A.~Joshi, and T.~Finin, ``A knowledge-based approach to
  intrusion detection modeling,'' in \emph{2012 IEEE Symposium on Security and
  Privacy Workshops}.\hskip 1em plus 0.5em minus 0.4em\relax IEEE, 2012, pp.
  75--81.

\bibitem{moscovich2020autosploit}
N.~Moscovich, R.~Bitton, Y.~Mallah, M.~Inokuchi, T.~Yagyu, M.~Kalech,
  Y.~Elovici, and A.~Shabtai, ``Autosploit: A fully automated framework for
  evaluating the exploitability of security vulnerabilities,'' \emph{arXiv
  preprint arXiv:2007.00059}, 2020.

\bibitem{mulwad2011extracting}
V.~Mulwad, W.~Li, A.~Joshi, T.~Finin, and K.~Viswanathan, ``Extracting
  information about security vulnerabilities from web text,'' in
  \emph{Proceedings of the 2011 IEEE/WIC/ACM International Conferences on Web
  Intelligence and Intelligent Agent Technology-Volume 03}.\hskip 1em plus
  0.5em minus 0.4em\relax IEEE Computer Society, 2011, pp. 257--260.

\bibitem{ou2006scalable}
X.~Ou, W.~F. Boyer, and M.~A. McQueen, ``A scalable approach to attack graph
  generation,'' in \emph{Proceedings of the 13th ACM conference on Computer and
  communications security}, 2006, pp. 336--345.

\bibitem{ou2005mulval}
X.~Ou, S.~Govindavajhala, and A.~W. Appel, ``Mulval: A logic-based network
  security analyzer.'' in \emph{USENIX security symposium}, vol.~8.\hskip 1em
  plus 0.5em minus 0.4em\relax Baltimore, MD, 2005, pp. 113--128.

\bibitem{pennington2014glove}
J.~Pennington, R.~Socher, and C.~D. Manning, ``Glove: Global vectors for word
  representation,'' in \emph{Proceedings of the 2014 conference on empirical
  methods in natural language processing (EMNLP)}, 2014, pp. 1532--1543.

\bibitem{peters2018deep}
M.~E. Peters, M.~Neumann, M.~Iyyer, M.~Gardner, C.~Clark, K.~Lee, and
  L.~Zettlemoyer, ``Deep contextualized word representations,'' \emph{arXiv
  preprint arXiv:1802.05365}, 2018.

\bibitem{rumelhart1985learning}
D.~E. Rumelhart, G.~E. Hinton, and R.~J. Williams, ``Learning internal
  representations by error propagation,'' California Univ San Diego La Jolla
  Inst for Cognitive Science, Tech. Rep., 1985.

\bibitem{sharnagat2014named}
R.~Sharnagat, ``Named entity recognition: A literature survey,'' \emph{Center
  For Indian Language Technology}, 2014.

\bibitem{simran2019deep}
K.~Simran, S.~Sriram, R.~Vinayakumar, and K.~Soman, ``Deep learning approach
  for intelligent named entity recognition of cyber security,'' in
  \emph{International Symposium on Signal Processing and Intelligent
  Recognition Systems}.\hskip 1em plus 0.5em minus 0.4em\relax Springer, 2019,
  pp. 163--172.

\bibitem{speicher2018formally}
P.~{Speicher}, M.~{Steinmetz}, R.~{Künnemann}, M.~{Simeonovski},
  G.~{Pellegrino}, J.~{Hoffmann}, and M.~{Backes}, ``Formally reasoning about
  the cost and efficacy of securing the email infrastructure,'' in \emph{2018
  IEEE European Symposium on Security and Privacy (EuroS P)}, 2018, pp. 77--91.

\bibitem{speicher2018stackelberg}
P.~Speicher, M.~Steinmetz, M.~Backes, J.~Hoffmann, and R.~K{\"u}nnemann,
  ``Stackelberg planning: Towards effective leader-follower state space
  search,'' in \emph{Thirty-Second AAAI Conference on Artificial Intelligence},
  2018.

\bibitem{stan2019heuristic}
O.~Stan, R.~Bitton, M.~Ezrets, M.~Dadon, M.~Inokuchi, Y.~Ohta, T.~Yagyu,
  Y.~Elovici, and A.~Shabtai, ``Heuristic approach towards countermeasure
  selection using attack graphs,'' \emph{arXiv preprint arXiv:1906.10943},
  2019.

\bibitem{stan2019extending}
O.~Stan, R.~Bitton, M.~Ezrets, M.~Dadon, M.~Inokuchi, Y.~Ohta, Y.~Yamada,
  T.~Yagyu, Y.~Elovici, and A.~Shabtai, ``Extending attack graphs to represent
  cyber-attacks in communication protocols and modern it networks,''
  \emph{arXiv preprint arXiv:1906.09786}, 2019.

\bibitem{sutskever2013importance}
I.~Sutskever, J.~Martens, G.~Dahl, and G.~Hinton, ``On the importance of
  initialization and momentum in deep learning,'' in \emph{International
  conference on machine learning}, 2013, pp. 1139--1147.

\bibitem{DeepExploit}
I.~Takaesu, ``Deepexploit, from
  https://github.com/13o-bbr-bbq/machine\textunderscore learning\textunderscore
  security/tree/master/deepexploit,'' 2018.

\bibitem{tharwat2018classification}
A.~Tharwat, ``Classification assessment methods,'' \emph{Applied Computing and
  Informatics}, 2018.

\bibitem{weerawardhana2014automated}
S.~Weerawardhana, S.~Mukherjee, I.~Ray, and A.~Howe, ``Automated extraction of
  vulnerability information for home computer security,'' in
  \emph{International Symposium on Foundations and Practice of Security}.\hskip
  1em plus 0.5em minus 0.4em\relax Springer, 2014, pp. 356--366.

\bibitem{werbos1990backpropagation}
P.~J. Werbos, ``Backpropagation through time: what it does and how to do it,''
  \emph{Proceedings of the IEEE}, vol.~78, no.~10, pp. 1550--1560, 1990.

\bibitem{yadav2019survey}
V.~Yadav and S.~Bethard, ``A survey on recent advances in named entity
  recognition from deep learning models,'' \emph{arXiv preprint
  arXiv:1910.11470}, 2019.

\bibitem{zhang2011empirical}
S.~Zhang, D.~Caragea, and X.~Ou, ``An empirical study on using the national
  vulnerability database to predict software vulnerabilities,'' in
  \emph{International conference on database and expert systems
  applications}.\hskip 1em plus 0.5em minus 0.4em\relax Springer, 2011, pp.
  217--231.

\bibitem{zhou2002named}
G.~Zhou and J.~Su, ``Named entity recognition using an hmm-based chunk
  tagger,'' in \emph{Proceedings of the 40th Annual Meeting of the Association
  for Computational Linguistics}, 2002, pp. 473--480.

\bibitem{zhu2015aligning}
Y.~Zhu, R.~Kiros, R.~Zemel, R.~Salakhutdinov, R.~Urtasun, A.~Torralba, and
  S.~Fidler, ``Aligning books and movies: Towards story-like visual
  explanations by watching movies and reading books,'' in \emph{Proceedings of
  the IEEE international conference on computer vision}, 2015, pp. 19--27.

\end{thebibliography}

%\clearpage
\onecolumn
\label{sec:append}
\begin{figure}[t!]
    \centering
    \includegraphics[width=1.0\textwidth]{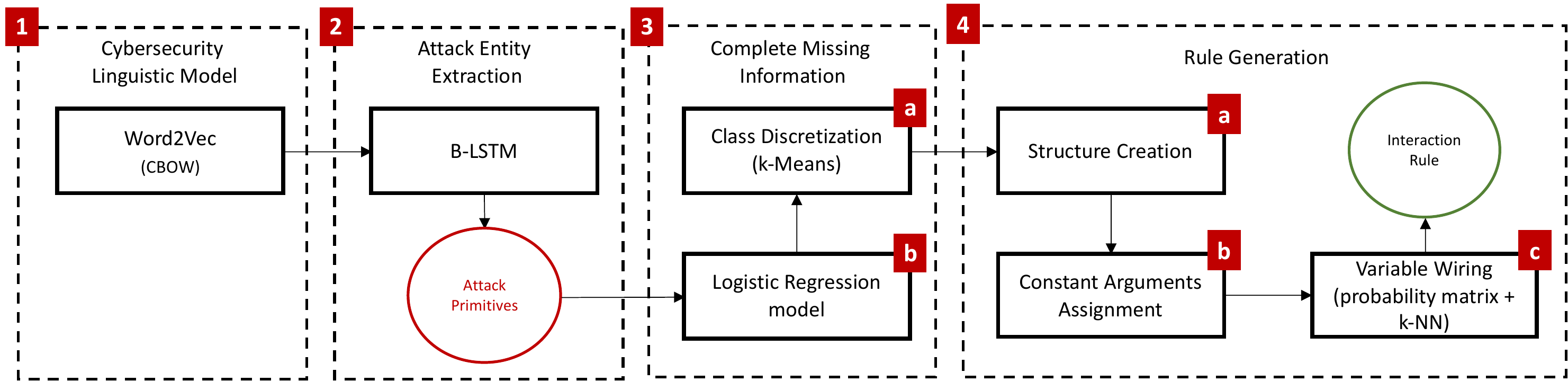}
    \caption{High level architecture of the proposed framework for modeling attacks from security vulnerabilities..}
    \label{fig:framework}
\end{figure}

\tiny
\begin{longtable}{|l|m{1.2cm}|m{1.2cm}|m{1.2cm}|l|m{1.2cm}|                m{0.1cm}|m{0.1cm}|m{0.1cm}|m{0.1cm}|m{0.1cm}|m{0.1cm}|m{0.1cm}|m{0.1cm}|m{0.1cm}|m{0.1cm}|m{0.1cm}|l|l|l|}
\hline
\multicolumn{1}{|c|}{\multirow{2}{*}{Work}}& 
\multicolumn{1}{c|}{\multirow{2}{*}{Dataset}}&
\multicolumn{1}{c|}{\multirow{2}{*}{Features}}&
\multicolumn{1}{c|}{\begin{tabular}[c]{@{}c@{}}Entities\\Extraction\end{tabular}} & 
\multicolumn{1}{c|}{\begin{tabular}[c]{@{}c@{}}Missing\\Values\end{tabular}} & 
\multicolumn{1}{c|}{\begin{tabular}[c]{@{}c@{}}Attack\\Modeling\end{tabular}} & 
\multicolumn{11}{c|}{Entities}& 
\multicolumn{3}{c|}{Reported Results}
% \multicolumn{1}{c|}{\begin{tabular}[c]{@{}c@{}}More\\Entities\end{tabular}}
\\ \cline{7-20}
&  &  &  &  &  & 
\rotatebox{90}{Impact} & 
\rotatebox{90}{Software} & 
\rotatebox{90}{OS} & 
\rotatebox{90}{Hardware} & 
\rotatebox{90}{Version} & 
\rotatebox{90}{Attack Action} &
\rotatebox{90}{Mean} & 
\rotatebox{90}{Privileges} & 
\rotatebox{90}{Port} & 
\rotatebox{90}{Attack Vector}& 
\rotatebox{90}{Network Protocol} & 
P & R & F1
% \multicolumn{1}{c|}{}
\\ \hline
\cite{li2019self} & 
APT reports and alienvault’ blogs with the mode BIO & 
Word embedding + Char embedding & 
BiLSTM + self attention & 
NO & 
NO & 
\ding{56} & \ding{52} & \ding{56} & \ding{56} & \ding{56} & \ding{52} & \ding{56} & \ding{56} & \ding{56} & \ding{56} & \ding{56} & 
0.86 & 0.83 & 0.84 \\
\hline
\cite{simran2019deep} & From \cite{bridges2013automatic}, CVE's descriptions from NVD  & Word embedding     & GRU + CNN + CRF   & NO     & NO     & \ding{56} & \ding{52} & \ding{56} & \ding{56} & \ding{52} & \ding{52} & \ding{56} & \ding{56} & \ding{56} & \ding{56} & \ding{56} &
0.9 & 0.96 & 0.93     \\
\hline
\cite{kim2020automatic} & 160 CTI reports (PDF files) &
BOC characterlevel feature representation (pre-trained Glove)+ hand crafted features & 
Bi‑LSTM ‑ CRF & 
NO & 
NO &
\ding{56} & \ding{56} & \ding{56} & \ding{56} & \ding{56} & \ding{56} & \ding{56} & \ding{56} & \ding{56} & \ding{56} & \ding{56} &
 & & 0.79  \\
\hline
\cite{gasmi2019cold} &
corpora of Bridges et al & 
Word embedding & 
Bi‑LSTM ‑ CRF & 
NO & 
ontology using the SWRL language &
\ding{56} & \ding{52} & \ding{52} & \ding{56} & \ding{52} & \ding{56} & \ding{56} & \ding{56} & \ding{56} & \ding{56} & \ding{56} & 
0.84 & 0.77 & 0.79 \\
\hline
\cite{jing2017augmenting} & 
540 CVEs descriptions (chosen by vulnerability type) & & 
Rule-Base, by punctuation and keywords & 
NO & 
Modeling to MuLVAl &
\ding{56} & \ding{52} & \ding{52} & \ding{56} & \ding{56} & \ding{56} & \ding{52} & \ding{52} & \ding{56} & \ding{52} & \ding{56} &
 & 0.88 & \\
\hline
\cite{lal2013information} &
30 security blogs, 240 CVE descriptions and 80 official security bulletins from Microsoft and Adobe & 
hand crafted features & 
CRF (Stanford NER) &
No & 
NO & 
\ding{52} & \ding{52} & \ding{52} & \ding{52} & \ding{52} & \ding{52} & \ding{56} & \ding{56} & \ding{56} & \ding{56} & \ding{56} & 
0.76 & 0.83 & 0.8 \\
\hline
\cite{weerawardhana2014automated} &
210 vulnerability descriptions from NVD & 
hand crafted features &
CRF (Stanford NER) &
NO &
NO & 
\ding{52} & \ding{52} & \ding{52} & \ding{56} & \ding{52} & \ding{56} & \ding{52} & \ding{56} & \ding{56} & \ding{56} & \ding{56} & 
0.54 & 0.58 & 0.55 \\
\hline
\cite{weerawardhana2014automated} & 
210 vulnerability descriptions from NVD & 
Part-Of-Speech & 
Rule-Base & 
NO &
NO & 
\ding{52} & \ding{52} & \ding{52} & \ding{56} & \ding{52} & \ding{56} & \ding{52} & \ding{56} & \ding{56} & \ding{56} & \ding{56} & 
0.73 & 0.82 & 0.71 \\
\hline
\cite{bridges2013automatic} & 
vulnerability descriptions from NVD & 
hand crafted features & 
Maximum Entropy Model &
NO&
NO& 
\ding{56} & \ding{52} & \ding{56} & \ding{56} & \ding{52} & \ding{52} & \ding{56} & \ding{56} & \ding{56} & \ding{56} & \ding{56} & 
0.98 & 0.99 & 0.99 \\
\hline
\cite{mulwad2011extracting}\ & 
107 vulnerability descriptions from NVD &  & 
OpenCalais and Wikitology & 
NO & 
Semantic Web language OWL & 
\ding{56} & \ding{52} & \ding{56} & \ding{56} & \ding{56} & \ding{56} & \ding{52} & \ding{56} & \ding{56} & \ding{56} & \ding{56} & 
& 0.9 & \\
\hline
\cite{aksu2018automated} & 
550 vulnerability descriptions from NVD & 
taxonomy-based & 
Rule-Based vs. RBF Networks vs. SVM vs. NEAT vs. MPL & 
NO & 
talk about attack graph but don’t talk about a method for modeling & 
\ding{56} & \ding{56} & \ding{56} & \ding{56} & \ding{56} & \ding{56} & \ding{56} & \ding{52} & \ding{56} & \ding{56} & \ding{56} &
 & 0.96 & \\
\hline
\cite{jones2015towards} & 
62 news articles, blogs, and updates & & 
boot - strapping & 
NO & 
NO & 
\ding{56} & \ding{52} & \ding{56} & \ding{56} & \ding{52} & \ding{56} & \ding{56} & \ding{56} & \ding{56} & \ding{56} & \ding{56} & 
0.8 & 0.24 &  
\\ \hline

{\begin{tabular}[c]{@{}c@{}}Our\\Work\end{tabular}} &
650 CVEs desciptions from NVD &
Word-Embeddings& 
B-LSTM & 
YES & 
Modeling to \mulval & 
\ding{52} & \ding{52} & \ding{52} & \ding{52} & \ding{52} & \ding{52} & \ding{52} & \ding{52} & \ding{52} & \ding{52} & \ding{52} & 
0.88 & 0.78 & 0.84  
\\ \hline
\caption{Summary of related works for attack entities extraction.\label{table-works}}
\end{longtable}

\twocolumn

\end{document}